\documentclass[fleqn,usenatbib]{mnras}

\usepackage{newtxtext,newtxmath}

\usepackage[T1]{fontenc}
\DeclareRobustCommand{\VAN}[3]{#2}
\let\VANthebibliography\thebibliography
\def\thebibliography{\DeclareRobustCommand{\VAN}[3]{##3}\VANthebibliography}

\usepackage{graphicx}	
\usepackage{amsmath}	
\usepackage{comment}
\usepackage{ulem}
\usepackage{pdflscape}	
\usepackage{lineno}

\defcitealias{Chen25_mockalgorithm}{C25}

\newcommand{\masyr}{mas\,yr$^{-1}$}
\newcommand{\kms}{km\,s$^{-1}$}
\newcommand{\kmskpc}{km\,s$^{-1}$\,kpc$^{-1}$}
\newcommand{\kmskpcGyr}{km\,s$^{-1}$\,kpc$^{-1}$\,Gyr$^{-1}$}
\newcommand{\lon}{$\phi_1$}
\newcommand{\lat}{$\phi_2$}
\newcommand{\pmone}{$\mu_{\phi_1}$}
\newcommand{\pmtwo}{$\mu_{\phi_2}$}

\title[Constraining the bar using the M92 stream]{Constraining the Galactic bar using the M92 stellar stream}

\author[A. Bystr\"om et al.]{Amanda Bystr\"om$^{1}$\thanks{E-mail: Amanda.Bystrom@ed.ac.uk},
Sergey E. Koposov$^{1,2}$,
Ting. S. Li$^{3}$,
Constance M. Rockosi$^{4}$,
Arjun Dey$^{5}$,
\newauthor
Guillaume F. Thomas$^{6,7}$,
Raymond G. Carlberg$^{3}$, 
Oleg Y. Gnedin$^{8}$,
Namitha Kizhuprakkat$^{9, 10}$,
\newauthor
Mika Lambert$^{4}$,
Nasser Mohammed$^{3}$,
Gustavo E. Medina$^{3}$,
Joan Najita$^{5}$,
Alexander H. Riley$^{11}$,
\newauthor
Nathan R. Sandford$^{3}$,
Leandro {Beraldo e Silva}$^{12}$,
Jessica N. Aguilar$^{13}$,
Steven Ahlen$^{14}$,
\newauthor
Davide Bianchi$^{15, 16}$,
David Brooks$^{17}$,
Todd Claybaugh$^{13}$,
Andrew P. Cooper$^{9}$,
Andrei Cuceu$^{13}$,
\newauthor
Axel de la Macorra$^{18}$,
Peter Doel$^{17}$,
Jaime E. Forero-Romero$^{19, 20}$,
Satya~{Gontcho A Gontcho}$^{21}$,
\newauthor
Gaston Gutierrez$^{22}$,
Richard Joyce$^{5}$,
Stephanie Juneau$^{5}$,
Anthony Kremin$^{13}$,
Martin Landriau$^{13}$,
\newauthor
Laurent Le~Guillou$^{23}$,
Aaron Meisner$^{5}$,
Ramon Miquel$^{24, 25}$,
Will J.~Percival$^{26, 27, 28}$,
Francisco Prada$^{29}$,
\newauthor
Ignasi P\'erez-R\`afols$^{30}$,
Graziano Rossi$^{31}$,
Eusebio Sanchez$^{32}$,
David Schlegel$^{13}$,
Joseph H. Silber$^{13}$,
\newauthor
Gregory Tarl\'{e}$^{33}$,
Benjamin A.~Weaver$^{5}$,
and Hu Zou$^{34}$
\\
Affiliations are listed at the end of the paper.
}

\date{Accepted XXX. Received YYY; in original form ZZZ}

\pubyear{\the\year{}}

\begin{document}
\label{firstpage}
\pagerange{\pageref{firstpage}--\pageref{lastpage}}
\maketitle

\begin{abstract}
Stellar streams are excellent probes of the gravitational potential in which they evolve. 
In the Milky Way (MW), globular cluster (GC) streams are routinely used to infer properties about time-dependent perturbations of the underlying potential. 
This implies that streams with Galactocentric radii small enough to be perturbed by the MW bar should offer constraints on it, such as its pattern speed, which currently has a wide range of values reported in the literature and is important when studying stellar kinematics. 
The GC M92 has a small pericentre and should be affected by the bar. 
It has a diffuse stellar stream, but confirming stream members has previously been hindered by a lack of spectroscopic data. 
In this paper, we use Dark Energy Spectroscopic Instrument (DESI) observations together with photometric and astrometric data to obtain spectroscopic members of the M92 stream for the first time. 
We identify a clear spatial distribution and gradients in distance moduli, proper motions, and radial velocities that confirm the stream's existence. 
We compare the observed stream to mock streams generated in different barred potentials and estimate the MW bar's pattern speed $\Omega = 29.1^{+0.7}_{-0.4}$ \kmskpc\ and $\dot \Omega = 0.7^{+3.5}_{-2.3}$ \kmskpcGyr. 
This is the first time a stellar stream is used to probabilistically infer these bar properties, and it opens up an exciting realm of inner Galactic potential characterisation using stellar streams.
\end{abstract}




\section{Introduction}

Stellar streams are elongated stellar structures that form when gravitationally bound dwarf galaxies or globular clusters (GCs) shed their constituent stars as they plunge into their host's gravitational potential. 
This makes streams excellent tracers of the underlying potential in which they evolve \citep{Johnston99_streamspotential, Ibata01_greatcirclestream, Helmi04_potential, Johnston05_potential, Fellhauer06_Sgrbifurcation, Law09_Sgrpotential, Koposov10_GD1, Bonaca14_Sgrpotential, Dierickx17_Sgrpotential, Malhan19_potential, Vasiliev21_tangoforthree, Ibata24_potential, Nibauer25_GD1potential}. 
In the Milky Way (MW), we know of over 120 streams \citep{Bonaca25_streamsreview}, out of which 80 are thought to be of GC origin \citep{Mateu23_galstreams}.
Compared to dwarf galaxy streams, GC streams are dynamically colder \citep{Li22_S5onedozenstreams} and have lower metallicity dispersions \citep{Li22_S5onedozenstreams, Geha26_GCDGs} due to the less efficient star formation history in GCs \citep{Leaman12_metallicitydispersions}.

Though it may appear like streams trace their progenitor's orbit, one single orbit is a poor description of all stars in a stream \citep{Sanders13_streamsdelineateorbits}.
This is because the host potential is complex \citep{Eyre11_streamsdonottraceorbit} and contains substructure and time-dependent perturbations that may influence the stream, on top of internal kinematics causing density variations along the stream such as epicyclic overdensities \citep{kupper_structure_2008, kupper_more_2012}.
These perturbations from the host potential can come from giant molecular clouds \citep{Amorisco16_GMCstreams}, 
GCs \citep{Doke22_GD1gapsGC, Ferrone25_GCstreamsimpact}, 
dark matter subhalos \citep{Johnston02_DMsubhaloes, Ibata02_DMsubhalostreams, Siegal-Gaskins08_DMsubhaloes, Carlberg09_DMsubhaloes, Yoon11_DMsubhaloes, Ngan14_DMsubhaloes, Erkal15_DMsubhalostreams, Bonaca19_DMsubhaloGD1, Li21_AtlasAliqaUma, Carlberg23_DMsubhaloes, Carlberg25_GD1subhalos}, 
the disc's spiral arms \citep{Banik19_Pal5substructure}, 
the Large Magellanic Cloud \citep{Koposov19_OC, Erkal19_OC, Lilleengen23_LMCOC, Brooks25_LMCstreams, Weerasooriya25_LMCstreams}, 
the Galactic bar \citep{Price-Whelan16_ophiuchusbar, Pearson17_Pal5bar, Thomas23_Hyadesbar}, 
or some combination of the above \citep{Erkal17_Pal5perturbations}.
Because streams are affected by these perturbations, observed streams can be used to constrain parameters of the substructure that causes them.

The fact that the MW contains a central bar has been known for a long time \citep{Dwek95_barexistence}, and we also have a good grasp of its morphology, such as the viewing angle and the radial extent \citep{Wegg15_baranglelength}.
There are different ages reported for the MW bar, but it may be older than 8 Gyr \citep[and references therein]{Sanders24_barage}.
What is still not well constrained is the rotation of the bar, the so-called pattern speed $\Omega$.
The range of values presented in the literature in recent years is large, ranging from 24 \kmskpc\ \citep{Horta25_barspeed} to 54 \kmskpc\ \citep{Ramos18_barspeed}, though most works are converging to values in the range $35-40$ \kmskpc\ (see the review by \citealt{Hunt25_MWreview} and their Fig. 10 for a more detailed comparison of published bar pattern speeds). 
The value of $\Omega$ will affect the dynamics of stars close enough to the bar to be perturbed by it and is thus important in kinematical modelling.

The bar pattern speed can be constrained by using the kinematics of stars in the inner galaxy, as a rotating bar will induce resonances which can be identified in kinematic space as overdensities \citep{Antoja14_HerculesOLR, Monari19_barresonance, Dillamore23_barresonance}.
This MW bar rotation should be slowing, as simulations predict that bars lose angular momentum to the disc and the halo \citep{Debattista98_barsloseAM, Debattista00_barsloseAM, Athanassoula03_bardeceleration, MartinezValpuesta06_bardeceleration}.
However, though the pattern speed should slow down on large timescales, it can also accelerate on shorter timescales \citep{Sellwood06_baracceleration} due to tidal interactions \citep{Lokas14_baracceleration}.
The bar's changing pattern speed needs to be taken into account for accurate modelling of kinematical substructures formed by the bar \citep{Dillamore25_barspeed}, as a slowing bar will leave different kinematical imprints than a rigidly rotating bar \citep{Fux01_bardeceleration}.
\citet{Chiba21_bardeceleration} measured the bar to be decelerating at the rate $\dot \Omega = -4.5$ \kmskpcGyr, using kinematic overdensities.
By introducing a slowing bar to match kinematic substructure, they were fully able to explain the morphology of the Hercules kinematic stream, and thus were able to resolve tension in the literature regarding what caused it.
This tension was between a fast and short bar with $\Omega > 50$ \kmskpc\ where Hercules consists of stars near the outer Lindblad resonance \citep{Antoja14_HerculesOLR, Fragkoudi19_HerculesOLR}, and a slow and long bar with $\Omega < 40$ \kmskpc, where Hercules stars are due to orbits trapped in the corotation resonance \citep{Perez-Villegas17_HerculesCR, DOnghia20_HerculesCR}.
Despite the importance of $\dot \Omega$ in bar models, there is still a paucity of bar deceleration constraints in the literature. 

Bar induced resonances affect not only field stars but also stellar streams as the resonances can create bifurcations and fanning features \citep{Yavetz23_streamresonances}.
There are not many examples of streams being affected by the bar in the literature, in part because many detected streams are located at large Galactocentric distances \citep{Mateu23_galstreams}, where the effects of the bar are negligible. 
One well-studied stream that may have been influenced by the bar is the GC stream Palomar 5.
The leading arm of Palomar 5 has a density truncation which was explained by \citet{Pearson17_Pal5bar} as being due to the bar, because the bar will exert different net torques on the stream stars depending on where they are in their orbits, which explains why the trailing arm is not truncated.
\citet{Erkal17_Pal5perturbations} saw similarly that the bar can induce asymmetries in the Palomar 5 stream even at orbital pericentres as large as 7 kpc. 

Another example of a bar-affected GC stream is Ophiuchus.
It has an old stellar population and is very short, and its extent indicates recent disruption which was difficult to explain considering its lack of a known progenitor \citep{Bernard14_ophiuchus, Sesar15_ophiuchus}.
Additionally, it has a spur feature extending from the main stream \citep{Caldwell20_Ophiuchusspur}.
To explain how Ophiuchus can have such a young dynamical age given its old stellar ages, as well as how some member stars end up far from the main stream, \citet{Price-Whelan16_ophiuchusbar} proposed that the bar induces chaos that truncates the stream.
\citet{Hattori16_ophiuchus} instead claim that the short stream length could be due to the progenitor repeatedly aligning with the bar at pericentre, causing a difference in torques exerted on stream stars at different points in their orbit. 
In a more recent paper by \citet{Yang25_Ophiuchusbar}, Ophiuchus was modelled with a decelerating bar, and using a changing $\Omega$, they were able to explain not just the short length and the fanning features of the stream but also the spur.
This shows the importance of considering a bar deceleration when simulating stellar streams in the presence of the bar.

The aforementioned papers have focused on qualitative matching of mock streams in barred potentials to observed streams. 
They use different bar pattern speeds that best reproduce their observations based on visual inspection, ranging from $\Omega = 60$ \kmskpc\ in the case of Palomar 5 \citep{Pearson17_Pal5bar} to $\Omega = 40$ \kmskpc\ for Ophiuchus \citep{Price-Whelan16_ophiuchusbar}.
The streams have not been used to go in the other direction by quantitatively constraining bar parameters.
However, though \citet{Yang25_Ophiuchusbar} are able to qualitatively reproduce the Ophiuchus stream features using $\dot \Omega = -5.5$ \kmskpcGyr\ and $\Omega = 35$ \kmskpc, they also design a metric to see which bar parameters from a grid of $\dot \Omega$ and $\Omega$ combinations quantitatively match their stream best, 
but note that using Ophiuchus to properly constrain these parameters is beyond the scope of their work.

M92 (or NGC 6341) is a metal-poor \citep{Geha26_GCDGs} GC on a very eccentric orbit, that was likely accreted onto the MW together with Gaia-Enceladus-Sausage \citep{Helmi18_GSE}.
Using \textit{Gaia} data, \citet{20Sollima_M92} identified an extended overdense region around M92 corresponding to its trailing arm. 
Using the matched filter-method on photometric data with metallicity information, \citet{Thomas20_M92} identified a similar faint, extended structure on the sky around M92, and confirmed its stream nature by tracer stars of different stellar populations as well as both a particle spray method and with a collisionless N-body simulation.
They found that the stream is only 500 Myr old.
Using only \textit{Gaia} data, \citet{Ibata21_STREAMFINDER} identified 84 stellar members of the M92 stream with the algorithm \texttt{STREAMFINDER} (though they carried out a follow-up spectroscopic programme that found radial velocities for a handful of stream members), and the follow-up work with a slightly altered algorithm and more spectroscopic follow-up observations measured radial velocities for 23 stream members.
Using the same \textit{Gaia} dataset, \citet{Chen25_StarStream} found 356 members using the algorithm \texttt{StarStream}.

The goal of this paper is to detect and characterise the M92 stellar stream using Dark Energy Spectroscopic Instrument (DESI) \citep{DESI22_DESI} spectroscopy together with Dark Energy Camera Legacy Survey (DECaLS) \citep{Dey19_DESILegacy} photometry and \textit{Gaia} \citep{Gaia23_GaiaDR3} astrometry. 
We then compare the resulting stream to mock streams in different barred MW potentials to constrain the Galactic bar's pattern speed and its rate of change.
For the first time, this is done with a probabilistic method using a stellar stream.
The paper is organised as follows.
In Sec. \ref{sec:data}, we present the DESI spectroscopic and the DECaLS photometric datasets used in this paper and in Sec. \ref{sec:streamid}, we show the methods used on these datasets to isolate stream members.
In Sec. \ref{sec:stream_modelling} we model the resulting stream and in Sec. \ref{sec:mockstreams} we present the methods to generate mock streams in different barred potentials, and how they are compared to the observed stream to get bar parameters.
The results are discussed in Sec. \ref{sec:discussion}.
Finally, a summary is given in Sec. \ref{sec:conclusions}.

\section{Data}\label{sec:data}

In this work, we rely on the combination of three distinct data sets that include \textit{Gaia} astrometry, DECaLS photometry, and DESI spectroscopy, which are restricted to the region within 6 deg of the M92 GC.
The DESI spectroscopic sample is described in Sec. \ref{sec:data_DESIspec} and the DECaLS sample in Sec. \ref{sec:data_DECaLSphot}.

Throughout the paper, we make use of known properties of the M92 GC, and these are presented in Table \ref{tab:progenitorinitialconds}.
We also adopt a stream-centric rotated-spherical coordinate system consisting of stream longitudes $\phi_1$ and stream latitudes $\phi_2$ \citep{Koposov10_GD1}.
The pole is at $(\alpha_p, \delta_p) = (93.72, 45.95)$ deg and the zero-point for $\phi_1$ is at the progenitor position at $\alpha_0 = 259.28$ deg.
We use this coordinate system to represent the stream's sky coordinates and proper motions in the paper from now on.

\begin{table}
	\centering
	\caption{
            Properties of the GC M92 used in this work.
            }
        \label{tab:progenitorinitialconds}
	\begin{tabular}{lll} 
            \hline
            Parameter & Value & Source \\
            \hline
            $\alpha$ [deg] & 259.28 & \citep{Baumgardt19_GCorbitalparams} \\
		      $\delta$ [deg] & 43.14 & \citep{Baumgardt19_GCorbitalparams} \\ 
            Distance [kpc] & 8.3 & \citep{Carney92_M92distance} \\
		      $\mu_\alpha$ [\masyr] & $-4.935$ & \citep{Baumgardt19_GCorbitalparams} \\ 
            $\mu_\delta$ [\masyr] & $-0.625$ & \citep{Baumgardt19_GCorbitalparams} \\
		      $v_R$ [\kms] & $-120.48$ & \citep{Baumgardt19_GCorbitalparams} \\
            $M$ [M$_\odot$] & $2.73 \cdot 10^5$ & \citep{Baumgardt18_M92params}\footnotemark \\
            $r_\mathrm{half-mass}$ [pc] & 3.59 & \citep{Baumgardt18_M92params}\footnotemark[\value{footnote}] \\
		\hline
	\end{tabular}
\end{table}
\footnotetext{We use the updated parameter values found at: \url{https://people.smp.uq.edu.au/HolgerBaumgardt/globular/fits/ngc6341.html}.}

\subsection{Spectroscopic DESI sample}\label{sec:data_DESIspec}

\subsubsection{The DESI spectroscopic survey}

The Dark Energy Spectroscopic Instrument (DESI) is a spectroscopic surveyor operating on the Mayall 4-meter telescope at Kitt Peak National Observatory \citep{DESI22_DESI}.
DESI uses 5,000 robotic fibre positioners over a $3.2$ deg-diameter field of view to obtain optical spectra for several targets simultaneously \citep{DESI16_instrumentdesign, Miller24_opticalcorrector, Poppett24_DESIfibers}, and is conducting an eight-year survey of about 17,000 square deg of the sky which began in May 2021 \citep{Schlafly23_DESIsurvey}.
The final survey will lead to 63 million galaxies and quasars \citep{Guy23_DESIdataprocessing} and the upcoming DESI second data release will include 12 million stars \citep{Koposov25_DESIDR1}.
DESI's first data release (DR1, \citealt{DESI26_DESIDR1}), which includes spectra taken between May 2021 through June 2022 for more than 18 million unique targets, is now public\footnote{\url{https://data.desi.lbl.gov/doc/releases/dr1/}}. 

The main mission of DESI is a galaxy and quasar redshift survey to measure baryonic acoustic oscillations and constrain the properties of dark energy by delivering the most precise measurement of the expansion history of the universe yet \citep{Levi13_snowmass}.
Early results from DESI have already set exciting constraints on the nature of dark energy \citep{DESI25_24VII, DESI25_25II}.
To achieve this, observations are made when the sky conditions are dark and transparency is good. 
However, during about 440 hours per year, the sky brightness is too high due to the moon and twilight, worsening seeing so that observations past redshift $0.6$ are difficult.
In these conditions, DESI makes observations for the Bright Time Survey, to which the Milky Way Survey (MWS) belongs, along with the Bright Galaxy Survey \citep{DESI16_surveydesign, Hahn23_BGS}.

\subsubsection{The DESI Milky Way Survey}\label{sec:data_DESIspec_MWS}

The main science goals of the DESI MWS are briefly to understand the chemical and dynamical evolution and star formation history of the MW’s thick disc and stellar halo, and to probe the MW's dark matter and accretion history \citep{Cooper23_MWSoverview}. 
This will be achieved by delivering radial velocities and chemical abundances for distant stars at high Galactic latitudes, targeting stars that are too faint for reliable velocities from other surveys such as \textit{Gaia}.
As part of DESI DR1, a so-called value added catalogue\footnote{\url{https://data.desi.lbl.gov/doc/releases/dr1/vac/mws/}} containing over 4 million stars with stellar parameter, abundance and radial velocity measurements was released \citep{Koposov25_DESIDR1}. 

The MWS has three main survey target categories, all observed during bright time, and out of these three the main sample is the biggest one.
It is split into \texttt{MAIN-BLUE}, \texttt{MAIN-RED} and \texttt{MAIN-BROAD}.
The targeting is based on DECaLS photometry and \textit{Gaia} astrometry, and their union is a magnitude-limited selection within $16<r_0<19$. 
Here, $r_0$ are extinction-corrected DECaLS $r$ magnitudes using the extinction coefficients in \citet{Schlafly11_dereddeningcoefs}, and from now on, all used DECaLS magnitudes are extinction-corrected and we continue to use the `$0$' subscript for clarity.
MWS observations also happen in dark time, which is when the main DESI cosmological survey is observed as dark time is executed under the best observing conditions.
See \citet{Cooper23_MWSoverview} for more details on these samples.
The MW backup programme complements the main survey by observing additional stars from the \textit{Gaia} catalogue in worse conditions, at brighter magnitudes, and at lower declination and Galactic latitude, while only using a small amount of DESI observation hours \citep{Dey25_backupprogramme}.
This MWS main survey bright programme sample forms the bulk of the DR1 stellar value added catalogue by contributing about 2.5 million stars, together with the MW backup and dark programmes which add about 1 million stars and 0.5 million stars respectively \citep{Koposov25_DESIDR1}.
In this work, we use all data, including the dark, bright and backup datasets described above, that will be included in the upcoming DESI DR2 which is currently scheduled for spring 2027 \citep{Koposov25_DESIDR1}.

Radial velocities are provided to all MWS targets via the radial velocity and stellar parameter fitting code RVSpecfit, which we will refer to as the RVS pipeline. 
The underlying idea behind the RVS pipeline is described in \citet{Koposov11_RVSapplication}, and it has been implemented in the Python package \texttt{RVSpecFit} \citep{Koposov19_RVS}, which is publicly available\footnote{\url{https://github.com/segasai/rvspecfit}}. 
This paper uses the most recent update of the RVS pipeline also used for the DR1 MWS value added catalogue \citep{Koposov25_DESIDR1}. 
On top of a star's radial velocity, the RVS pipeline also provides the stellar metallicity [Fe/H], and the uncertainties on these parameters, which are used in this work.
The RVS outputs are also used to create a 64 bit warning bitmask called $\texttt{RVS\_WARN}$, which consists of four individual bitmasks: \texttt{\{'CHISQ\_WARN': 1, 'RV\_WARN': 2, 'RVERR\_WARN': 4, 'PARAM\_WARN': 8\}}.
These flags are set if a given spectrum's difference in $\chi^2$ of the stellar model with respect to the continuum is larger than 50, if the fitted radial velocity is within 5 \kms\ of the velocity interval edges at [--1500, 1500] \kms, if the fitted radial velocity error is larger than 100 \kms, or if the derived metallicity or effective temperature is at the grid edges, respectively for each bitmask.
A good spectral fit which has none of these warning flags set have $\texttt{RVS\_WARN} = 0$.

\subsubsection{M92 stellar stream observations}\label{sec:data_DESIspec_M92targeting}

Special surveys are conducted in DESI to answer scientific questions not necessarily included in already defined surveys.
These have their own set of targeting strategies and criteria, and are conducted in parallel with the main survey.
These special surveys are sometimes referred to as tertiary programmes, and it is the terminology we will use from now on.
Tertiary programme targets are not processed by the \texttt{desitarget}\footnote{\url{https://github.com/desihub/desitarget/}} pipeline, which is used for main survey targets and keeps track of which sources will be targeted for DESI follow-up spectroscopy, and are instead handled only by the DESI \texttt{fiberassign}\footnote{\url{https://github.com/desihub/fiberassign}} code, which plans the location of DESI fiber positioners for every DESI target (Raichoor et al. in prep.).
This means that in the case of tertiary programme targets, the inputs to \texttt{fiberassign} does not come from \texttt{desitarget}, which is the case for all other targets.
See \citet{Myers23_DESItargetpipeline} for more information on DESI target selections and pipelines. 

\begin{figure}
 \includegraphics{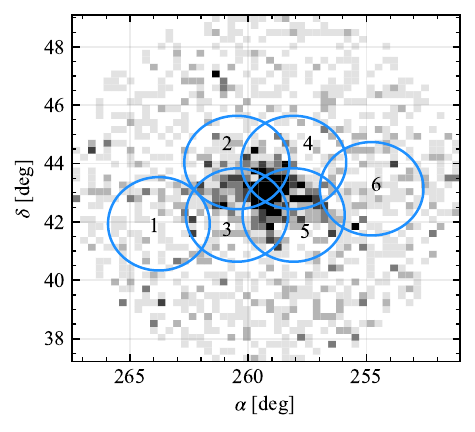}
 \caption{ 
          The density of the DECaLS photometric sample, after the initial stream selections described in Sec. \ref{sec:streamsel_initialsel} have been applied, shown in greyscale, and the DESI spectroscopic tertiary programme tiles are shown in blue.
          These DESI tiles are labelled from 1 to 6.
          The M92 GC is at the centre of the DESI tiles.
        }
 \label{fig:phot_sample}
 \end{figure}

In autumn 2023, DESI observed six tiles centred on the M92 GC and its outskirts during bright time in such a tertiary programme, with the purpose of mapping the M92 stellar stream. 
The tiles were designed to understand the requirements of future stream observations in preparation for the upcoming DESI survey extension. 
These DESI tertiary tiles are shown in blue in Fig. \ref{fig:phot_sample}, together with the DECaLS photometric sample as a two-dimensional histogram.
We return to the DECaLS sample in Sec. \ref{sec:data_DECaLSphot}.
In this paper, we analyse objects both from this tertiary programme and the main MWS sample described in Sec. \ref{sec:data_DESIspec_MWS}.
From now on we refer to the combined dataset as the spectroscopic sample.

The M92 tertiary programme target list is split into four categories that do not overlap, listed here in order of decreasing priority: central targets \texttt{CENTRAL}, proper motion and colour-magnitude diagram (CMD) selected targets \texttt{PMCMD}, proper motion selected targets \texttt{PM}, and filler targets \texttt{FILLER}.
The first three target categories have the same astrometric requirements. 
The \texttt{CENTRAL} targets, which have the highest priority, are within 0.2 deg of the progenitor GC on the sky, and the targeting is designed to avoid crowding issues in the dense cluster region by avoiding stars that are faint and have close neighbours.
The remaining three target categories are more than 0.2 deg away from the GC centre.
The \texttt{PMCMD} targets have colours within 0.075 mag of the progenitor CMD track, which the \texttt{PM} targets do not, and so the latter targets have an additional colour cut applied to them.
Both the \texttt{PMCMD} and \texttt{PM} targets are fainter than the \texttt{CENTRAL} targets, to allow for deeper observations of the fainter regions outside the progenitor.
The \texttt{FILLER} targets are those that do not belong to any other category.
Their targeting criteria are given in Table \ref{tab:target_criteria} and the priorities, given by the column \texttt{PRIORITY\_INIT}, are 8000 for the \texttt{CENTRAL} targets, 7000 for the \texttt{PMCMD} targets, 6000 for the \texttt{PM} targets and 5000 for the \texttt{FILLER} targets; see \citet{Schlafly23_DESIsurvey} for more information on DESI priority assignment.
The final set of spectroscopic observation targets is a union of these four targeting categories.

\begin{table}
	\centering
	\caption{
                Targeting criteria for the four M92 tertiary programme targets. 
                In the table, the target parameters are as follows: $\alpha$ and $\delta$ are right ascension and declination; $\mu_\alpha$ and $\mu_\delta$ are the \textit{Gaia} proper motions in right ascension and declination respectively and $\varepsilon_\mu$ the total proper motion error; $\varpi$ is \textit{Gaia} parallax and $\varepsilon_\varpi$ the parallax error; $G$ and $G_\text{RP}$ are \textit{Gaia} G-band and RP-band magnitudes; and $r_0$ and $g_0$ are DECaLS magnitudes.
                Except for when referring to magnitudes, a '0' subscript refers to the progenitor's observed properties given in Table \ref{tab:progenitorinitialconds}.
                $f(r_0)$ is a function of colour that is interpolated from the points given in Table \ref{tab:padCMD_values_targeting}.
                }
        \label{tab:target_criteria}
	\begin{tabular}{l} 
            \hline
            \texttt{CENTRAL}  \\
		  $\bullet \text{ } \sqrt{(\alpha - \alpha_0)^2 + (\delta - \delta_0)^2} < 0.2$ deg \\ 
            $\bullet \text{ } \sqrt{(\mu_\alpha - \mu_{\alpha, 0})^2 + (\mu_\delta - \mu_{\delta, 0})^2} < 2 + 2.5 \varepsilon_\mu$ mas yr$^{-1}$ \\
		  $\bullet \text{ } |\varpi - 1/d_0| < 0.05 + 2.5\varepsilon_\varpi$ mas \\ 
            $\bullet \text{ } 16 < G < 19$ mag \\
		  $\bullet$ no neighbours in the \textit{Gaia} catalogue within 1 arcsec \\ 
		\hline
            \texttt{PMCMD}  \\
		  $\bullet \text{ } \sqrt{(\alpha - \alpha_0)^2 + (\delta - \delta_0)^2} \geq 0.2$ deg \\ 
            $\bullet \text{ } \sqrt{(\mu_\alpha - \mu_{\alpha, 0})^2 + (\mu_\delta - \mu_{\delta, 0})^2} < 2 + 2.5 \varepsilon_\mu$ mas yr$^{-1}$ \\
		  $\bullet \text{ } |\varpi - 1/d_0| < 0.05 + 2.5\varepsilon_\varpi$ mas \\ 
            $\bullet \text{ } 16 < G < 21$ mag \\
		  $\bullet \text{ } |f(r_0) - (g_0-r_0)| < 0.075$ mag \\ 
		\hline
            \texttt{PM} \\
		  $\bullet \text{ } \sqrt{(\alpha - \alpha_0)^2 + (\delta - \delta_0)^2} \geq 0.2$ deg \\ 
            $\bullet \text{ } \sqrt{(\mu_\alpha - \mu_{\alpha, 0})^2 + (\mu_\delta - \mu_{\delta, 0})^2} < 2 + 2.5 \varepsilon_\mu$ mas yr$^{-1}$ \\
		  $\bullet \text{ } |\varpi - 1/d_0| < 0.05 + 2.5\varepsilon_\varpi$ mas \\ 
            $\bullet \text{ } 16 < G < 21$ mag \\
		  $\bullet \text{ } |f(r_0) - (g_0-r_0)| \geq 0.075$ mag \\ 
            $\bullet \text{ } G - G_\text{RP} < 0.8$ \\
		\hline
            \texttt{FILLER}  \\
		  $\bullet \text{ } \sqrt{(\alpha - \alpha_0)^2 + (\delta - \delta_0)^2} \geq 0.2$ deg \\ 
            $\bullet \text{ } 16 < G < 20$ mag \\
		  $\bullet$ do not belong to any of the previous categories \\ 
		\hline
	\end{tabular}
\end{table}

\begin{table}
	\centering
	\caption{
                The points in colour $g_0-r_0$ and apparent magnitude $r_0$ along the GC CMD that are interpolated to create the spline function $f(r_0)$ in the M92 DESI targeting.
                }
        \label{tab:padCMD_values_targeting}
	\begin{tabular}{cc} 
            \hline
            $g_0-r_0$ [mag] & $r_0$ [mag] \\
		\hline
		  0.57 & 21.0 \\ 
            0.38 & 20.0 \\
		  0.24 & 19.0 \\ 
            0.23 & 18.1 \\
		  0.43 & 17.3 \\ 
            0.55 & 15.0 \\
		\hline
	\end{tabular}
\end{table}

Observations of this tertiary programme consisted of 30 observations in total, as each of the six tiles were observed five times in a four-point dithering pattern.
This pattern was achieved by completing one pointing at the tile centre, and four additional pointings that were offset to the right and left of the centre by 0.065 deg and above and below the centre by 0.048 deg.
The tiles have been labelled in Fig. \ref{fig:phot_sample}, and the total effective exposure times \citep{Schlafly23_DESIsurvey} across all dithers are 1459.8 seconds for the first, 1583.1 seconds for the second, 1671.1 seconds for the third, 1509.1 seconds for the fourth, 1358.3 seconds for the fifth, and 1530.3 seconds for the sixth tile.
As a reference, the effective exposure time for a tile in a standard MWS observation is $\sim180$ seconds. 
The tertiary programme therefore reaches significantly greater depth, enabling sufficient S/N for fainter stars down to $G = 21$ mag, as shown in Table \ref{tab:target_criteria}.
The objects observed in this tertiary programme will be released in the upcoming DESI DR2, together with the main and backup MWS objects which are also used in this work.
Therefore, this paper studies all objects that will be included in DR2 (but we focus on those that are within six deg of the M92 GC).

Like the main MWS sample, the tertiary programme objects are processed through the RVS pipeline and have stellar parameters and radial velocities.
To the set of stars included in our combined DESI spectroscopic sample, we apply a set of quality cuts.
The first is the RVS quality flag $\texttt{RVS\_WARN} = 0$ which we apply to all candidates.
Additionally to these RVS quality flags, all objects that are flagged as quasars by the template-fitting code Redrock (Bailey et al. in prep.) are removed by requiring \texttt{RR\_SPECTYPE != QSO}, ensuring that our sample consists only of stars.
We also require that if a target has been observed more than once, we choose the best observation with \texttt{PRIMARY = True}\footnote{For a tutorial on how to use the MWS value added catalogue, including applying these quality flags, see \url{https://github.com/desimilkyway/dr1_tutorials}}.
To all stars' radial velocity errors, we add a systematic error floor of 0.9 \kms\ \citep{Koposov25_DESIDR1}.
After additionally applying the stream selection criteria derived in this work (see Sec. \ref{sec:streamsel_sky}), the spectroscopic sample is dominated by the tertiary programme stars.

Once the paper is accepted for publication, all 72,147 objects observed as part of this tertiary programme will be released with an accompanying tutorial that shows how to recreate the stream selection and figures of this paper.

\subsection{Photometric DECaLS sample}\label{sec:data_DECaLSphot}

It is useful to supplement the DESI spectroscopic sample with a larger photometric sample that has fewer selections and larger completeness, to map the stream with purely astrometric and photometric selections.
This sample is a combination of Dark Energy Camera Legacy Survey (DECaLS) and \textit{Gaia} data.
From the DECaLS \citep{Dey19_DESILegacy} DR9\footnote{\url{https://www.legacysurvey.org/dr9/}}, we query all objects that are within six deg of the progenitor GC M92.
The DECaLS sample is then crossmatched with \textit{Gaia} DR3 \citep{Gaia23_GaiaDR3}, requiring a maximum sky separation of 0.5 arcsec.
From this set of objects with DECaLS photometry and \textit{Gaia} astrometry, we only keep objects with \textit{Gaia} magnitudes $16 < G < 21$ mag. 
This creates our final set of objects, which we from now on refer to as the photometric sample.
The photometric sample is shown as the two-dimensional histogram in Fig. \ref{fig:phot_sample}. 
In the figure, we show the sample after an initial stream selection has been applied, which is described in Sec. \ref{sec:streamsel_initialsel}, to show the rough stream extent next to the DESI tiles, which are shown in blue.

\section{Stream member selection}\label{sec:streamid}

The following sections outline how we identify M92 stellar stream members.
First, we utilise only the photometric sample as it contains more stars than the spectroscopic sample.
We first apply an initial, broad stream selection to the photometric sample as outlined in Sec. \ref{sec:streamsel_initialsel} to identify the approximate stream extent.
In Sec. \ref{sec:streamsel_DMgrad} we find the distance modulus gradient of the stream, which we apply to the photometric sample in Sec. \ref{sec:streamsel_CMDsel} to get a cleaner CMD, which we design our CMD cuts around.
In Sec. \ref{sec:streamsel_pmtracks} we show how we select stars around the proper motion tracks of the stream using the photometric sample.
These astrometric and photometric selections are applied to the spectroscopic sample in Sec. \ref{sec:streamsel_fehvrad}, which allows us to create selections in metallicity and radial velocity.
Finally we summarise these selections and show the resulting sky morphology in Sec. \ref{sec:streamsel_sky}.

\subsection{Initial selection}\label{sec:streamsel_initialsel}

The starting point for selecting the M92 stellar stream is to apply cuts to the photometric sample using astrometry and photometry.
These initial cuts will allow us to isolate key features of the stream that we will utilise to refine our selections.
They are similar to some of the target selection criteria for DESI spectroscopic observations of M92 (see Sec. \ref{sec:data_DESIspec_M92targeting}), but are more conservative.

The first selection is in parallax $\varpi$, where we select stars in a heliocentric annulus centred on the distance of the progenitor GC at $d_0$ kpc (see Table \ref{tab:progenitorinitialconds}) taking the parallax error $\varepsilon_\varpi$ into account:

\begin{equation}\label{eq:parallax_sel}
    |\varpi - 1/d_0| < 0.1 + 2.5\varepsilon_\varpi
\end{equation}

\noindent where the padding values of 0.1 and 2.5 mas have been determined empirically.

The second selection is in proper motions in right ascension $\mu_\alpha$ and declination $\mu_\delta$.
We select stars in a circle centred on the GC proper motions presented in Table \ref{tab:progenitorinitialconds}, denoted by $\mu_{\alpha, 0}$ and $\mu_{\delta, 0}$, taking into account the sample stars' combined proper motion error $\varepsilon_\mu$:

\begin{equation}\label{eq:pm_padsel}
    \sqrt{(\mu_\alpha - \mu_{\alpha, 0})^2 + (\mu_\delta - \mu_{\delta, 0})^2} < 0.5 + \varepsilon_\mu
\end{equation}

\noindent where again the padding of 0.5 \masyr\ has been found empirically.

The third selection is done around the CMD distribution of the progenitor GC.
To identify the CMD selection area, we apply the parallax cut in Eq. \ref{eq:parallax_sel} and the proper motion cut in Eq. \ref{eq:pm_padsel}, as well as a sky selection centred on the GC, where all stars within 0.5 deg and outside 0.15 deg of the GC are kept.
This sky selection is applied to reveal the GC CMD only, so we know where to place our selection region, but is not kept for any subsequent steps.
Using the resulting CMD, a set of points in colour $(g_0-r_0)$ and apparent magnitude $r_0$ along the GC CMD are chosen by eye, and at this step, we ignore the blue horizontal branch (BHB) part of the CMD.
These points are given in Table \ref{tab:padCMD_values}. 
We interpolate between these points to create a colour track as a function of magnitude $f(r_0)$ (note that this function is different from that for the targeting criteria given in Table \ref{tab:target_criteria}, which interpolates the values in Table \ref{tab:padCMD_values_targeting}).
Then we select all stars within $0.055$ mag of this spline. 
The CMD selection is then defined as:

\begin{equation}\label{eq:CMD_padsel}
    |f(r_0) - (g_0-r_0)| < 0.055
\end{equation}

\begin{figure*}
 \includegraphics{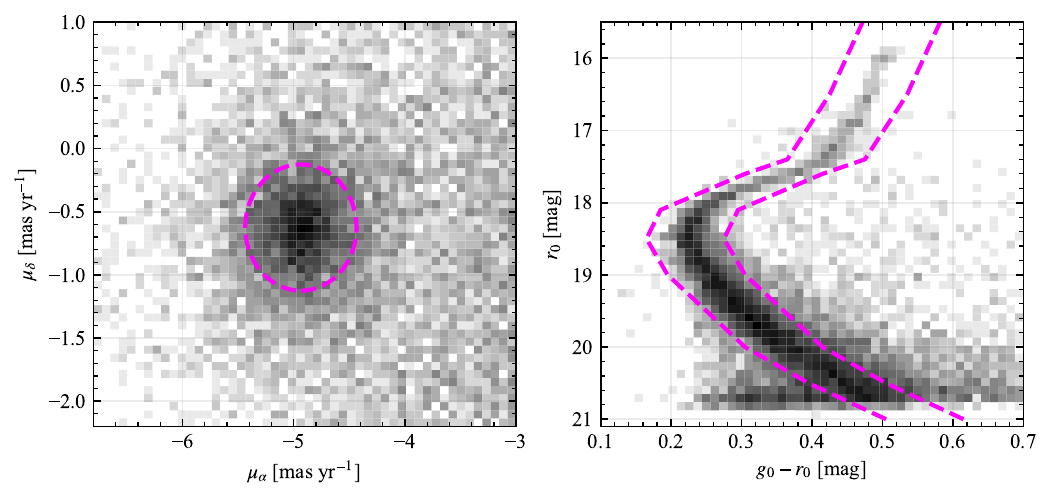}
 \caption{ 
          The initial selection of M92 stream stars.
          \textit{Left:} The distribution of \textit{Gaia} proper motions of the photometric sample, after a cut in parallax, Eq. \ref{eq:parallax_sel}, and CMD (shown on the right, Eq. \ref{eq:CMD_padsel}), has been applied. 
          The magenta dashed circle shows the cut in proper motion with a radius of 0.5 \masyr, as given in Eq. \ref{eq:pm_padsel}.
          \textit{Right:} The CMD of the photometric sample, after a cut in parallax, Eq. \ref{eq:parallax_sel}, and proper motions (shown on the left, Eq. \ref{eq:pm_padsel}), has been applied.
          The magenta dashed lines outline the region of CMD selection as defined by Eq. \ref{eq:CMD_padsel}.
        }
 \label{fig:pad_initialsel}
 \end{figure*}

The astrometric selections given in Eqs. \ref{eq:parallax_sel} and \ref{eq:pm_padsel} and the CMD region selected by Eq. \ref{eq:CMD_padsel}, applied to the photometric sample, are  shown in Fig. \ref{fig:pad_initialsel}.
The panel on the left shows the sample after the parallax selection and the CMD selection have been applied, with the selection circle of 0.1 \masyr in proper motion shown in magenta (not depicting the error range). 
The panel on the right shows the CMD of the sample, with the parallax and proper motion selections applied, and the CMD selection region outlined again in dashed magenta lines. 
These panels clearly show the overdensity of M92 stars in proper motion and CMD space. 
The CMD padding value of 0.055 mag in Eq. \ref{eq:CMD_padsel} is motivated by the thin CMD seen in the left-hand panel.

\begin{table}
	\centering
	\caption{
                The points in colour $g_0-r_0$ and apparent magnitude $r_0$ along the GC CMD that are interpolated to create the spline function $f(r_0)$ in Eq. \ref{eq:CMD_padsel}.
                }
        \label{tab:padCMD_values}
	\begin{tabular}{cc} 
            \hline
            $g_0-r_0$ [mag] & $r_0$ [mag] \\
		\hline
		  0.53 & 21.0 \\ 
            0.36 & 20.0 \\
		  0.26 & 19.0 \\ 
            0.23 & 18.5 \\
		  0.24 & 18.1 \\ 
            0.36 & 17.6 \\
		  0.42 & 17.4 \\ 
            0.55 & 15.0 \\
		\hline
	\end{tabular}
\end{table}

\subsection{Distance modulus track}\label{sec:streamsel_DMgrad}

Now that we have our initial stream selections, we wish to create a cleaner CMD to select stream stars from.
Our aim now is to identify the distance modulus distribution as a function of $\phi_1$ in our data, using well-established distance tracers, which then can be applied to the sample CMD to more accurately select stream stars in colour and absolute magnitude, as this correction will remove vertical smearing in the CMD induced by distance differences along the stream \citep{deBoer18_GD1, Li21_AtlasAliqaUma}.

The distance tracers we use are subgiant branch (SGB) stars.
We also search for BHB stars by selecting stars based on their surface gravities and effective temperatures following \citet{Bystroem25_LMC}, but do not find any close to the stream track.
SGB candidates are identified by selecting all stars in the photometric sample that have astrometry in agreement with the selections in Eqs. \ref{eq:parallax_sel} and \ref{eq:pm_padsel}, fall within the colour range $0.27 < (g_0 - r_0) < 0.40$, and have sky coordinates in the range:

\begin{equation}\label{eq:streamtrack}
    |\phi_2 - (0.041 \phi_1 - 0.176 \phi_1^2 +0.001 \phi_1^3)| < 1.3,
\end{equation}

\noindent where the polynomial is an approximation of the M92 stream track defined by \citet{Thomas20_M92}.
For the SGB candidates that pass the above cuts, we compute their absolute magnitudes $M_r$ assuming that they are at the GC distance of 8.3 kpc.
From the resulting colour-absolute magnitude diagram (CAMD), we choose stars along this sequence to interpolate a relationship of $M_r$ as a function of colour $(g_0-r_0)$ valid for SGB stars: 

\begin{equation}\label{eq:SGBabsmag}
    M_r = 5.75 -14.88(g_0-r_0) + 27.77(g_0-r_0)^2 -20.52(g_0-r_0)^3
\end{equation}

\begin{figure}
\includegraphics{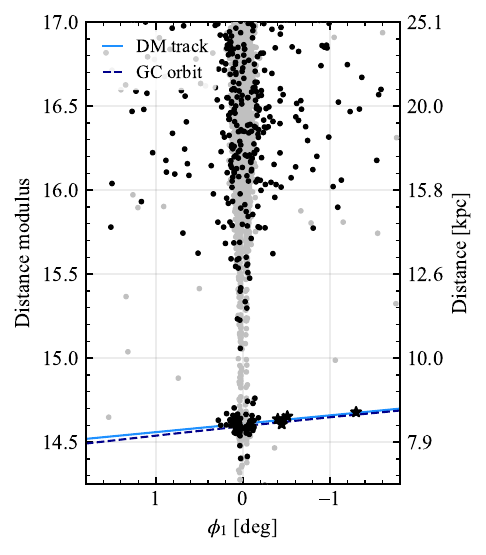}
 \caption{ 
          The distance modulus distribution computed using Eq. \ref{eq:SGBabsmag} as a function of \lon\ for all SGB candidate stars (grey points) and the spectroscopically selected stream members (black points), with the corresponding distance given on the right axis.
          The distance modulus track given in Eq. \ref{eq:DMgradient} is shown as a blue line and the integrated GC orbit is given as a dark blue dashed line.
          The five stars used for interpolating the distance modulus track are shown with star markers.
          }
 \label{fig:distgradient}
 \end{figure}

\noindent This is used with the SGB candidates' $r_0$ magnitude to get their distance moduli.
The resulting distance modulus distribution as a function of \lon\ is shown as grey points in Fig. \ref{fig:distgradient}.
This distribution peaks around $\phi_1 = 0$ deg because the SGB candidates are dominated by stars inside the GC.
The vertical distribution is due to most of the stars within our SGB colour range being main sequence stars, and when they are assumed to be SGB stars, their distance moduli become overestimated.

To interpolate a distance modulus track, we apply a spectroscopic selection to the SGB star candidates to choose stars in the stellar stream.
This consists of a metallicity and a radial velocity cut\footnote{The radial velocity cut used here is not the final one we derive in Sec. \ref{sec:streamsel_fehvrad}, but a cruder one that allows us to efficiently select tentative spectroscopic stream members.} defined as:
\begin{itemize}
\item $-3.5 < \mathrm{[Fe/H] < -1.5}$ dex, 
\item $|v_R - v_{R,0}| < 20 + 5 \varepsilon_{v_R}$
\end{itemize}
where [Fe/H] is the stellar metallicity, $v_R$ is the stellar radial velocity and $\varepsilon_{v_R}$ the corresponding errors, and $v_{R,0}$ is the mean radial velocity of the cluster given in Table \ref{tab:progenitorinitialconds}.
The resulting distribution of spectroscopically selected stream-like SGB candidates is shown as black points in Fig. \ref{fig:distgradient}.
Among these stars, five stand out at $\phi_1 < 0$ deg, as they have distance moduli similar to the GC, which means that they are true SGB stars, and the spectroscopic selection ensures they also belong to the stream.
They are distributed along a slope in distance modulus.
To confirm this, we integrate the progenitor orbit forward and backward in the potential \texttt{MilkyWayPotential2022} \citep{Eilers19_MWcircvelcurve, Darragh-Ford23_verticalspiral} implemented in \texttt{Gala} \citep{Price-Whelan17_gala, price-whelan_galaversion}. 
These five stars lie on top of this progenitor orbit, which we interpret as them being real SGB stream stars, and that the stream indeed has a distance modulus gradient.
They are shown with star markers in Fig. \ref{fig:distgradient} to highlight their position relative to this integrated orbit.
From these five stars, we interpolate the following distance modulus track:

\begin{equation}\label{eq:DMgradient}
    \mathrm{DM}_0 (\phi_1) = -0.05 \phi_1 + 14.61
\end{equation}

\noindent This track is plotted as a blue line in Fig. \ref{fig:distgradient}.
The integrated GC orbit is also shown, as a dark blue dashed line, which clearly aligns very well with the interpolated DM track.
It is used to compute the distance modulus of all stars in the photometric sample before making the CMD selection, to correct for any vertical smearing that this gradient would result in if uncorrected.

\subsection{Colour-absolute magnitude diagram selection}\label{sec:streamsel_CMDsel}

\begin{figure}
 \includegraphics{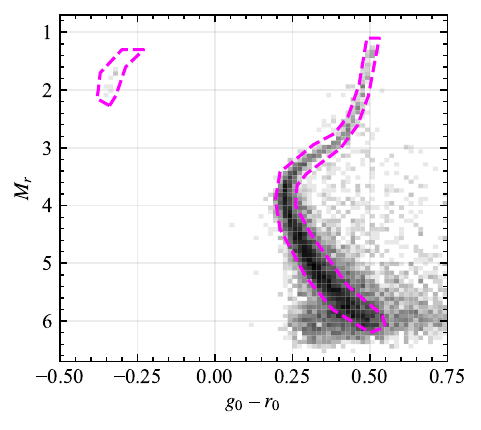}
 \caption{ 
          The CAMD of all stream star candidates, selected using the astrometric criteria given in Eqs. \ref{eq:parallax_sel} and \ref{eq:pm_padsel} and their distances corrected using Eq. \ref{eq:DMgradient}, shown as a two-dimensional histogram.
          The overlaid dashed magenta box shows the CMD selection applied to the stream stars.
          }
 \label{fig:CMD_sel}
\end{figure}

The next step is to isolate the stream stars in CAMD space, making use of the distance modulus track given in Eq. \ref{eq:DMgradient}.
We select stars around the progenitor GC in right ascension and declination, because the stream stars will share the same CMD properties as the GC members, but this bound GC selection allows for a cleaner view of the CMD.
We select all stars outside 0.15 deg of the GC and all stars within 0.5 deg of the GC.
Then the astrometric selections given in Eqs. \ref{eq:parallax_sel} and \ref{eq:pm_padsel} are applied.
We apply the distance modulus track given in Eq. \ref{eq:DMgradient} to all GC stars, so that we can compute their absolute magnitudes $M_r$ from their colours $r_0$.
Around the resulting GC CAMD, we hand-draw a selection box that includes the horizontal branch.
The selection box is shown in dashed magenta lines in Fig. \ref{fig:CMD_sel}, where the horizontal branch is the box at the upper left of the figure.
The two-dimensional histogram in the background shows the CAMD of the photometric sample after applying the astrometric selections in Eqs. \ref{eq:parallax_sel} and \ref{eq:pm_padsel}, and after computing absolute magnitudes using Eq. \ref{eq:DMgradient}.
A clear CAMD distribution can be seen within the selection region.

\subsection{Proper motion track selections}\label{sec:streamsel_pmtracks}

The proper motion selection given by Eq. \ref{eq:pm_padsel} only selects stars in a small circle around the progenitor. 
If the stream has a gradient in proper motion with $\phi_1$ it would mean that stream members with large $|\phi_1|$ have significantly different proper motions from the progenitor, causing the selection in Eq. \ref{eq:pm_padsel} to truncate the stream.
We thus now turn to fitting the stream's proper motion tracks in both the $\phi_1$ and $\phi_2$ directions, $\mu_{\phi_1}$ and $\mu_{\phi_2}$, so that we can make reliable selections in proper motions along the entire stream.

\begin{figure*}
 \includegraphics{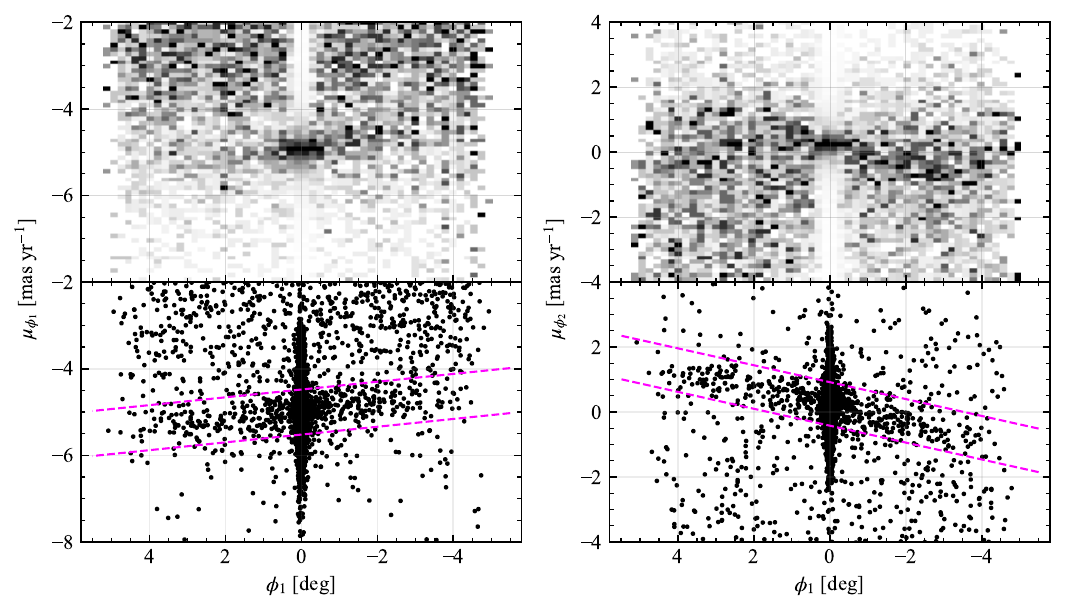}
 \caption{
          \textit{Top row:} The distribution in proper motions as a function of stream longitude $\phi_1$, along the stream longitude ($\mu_{\phi_1}$, top left) and stream latitude ($\mu_{\phi_2}$, top right), after the selections in Eqs. \ref{eq:parallax_sel} and \ref{eq:streamtrack} as well as the CMD selection have been applied.
          The two-dimensional histograms have been column-normalised.
          \textit{Bottom row:} Same as the top row, but for when stars inside the dashed magenta lines in the other proper motion direction have been selected. 
          The selection windows correspond to two times the proper motion dispersion in each direction, respectively.
        }
 \label{fig:pmtracks}
 \end{figure*}

To start with, all previous selections are applied to the photometric sample: the parallax selection in Eq. \ref{eq:parallax_sel} and the CMD selection shown in Fig. \ref{fig:CMD_sel}, as well as the sky selection in Eq. \ref{eq:streamtrack}. 
The column-normalised distributions of the two proper motion directions $\mu_{\phi_1}$ and $\mu_{\phi_2}$ as a function of $\phi_1$ are shown in the top row of Fig. \ref{fig:pmtracks}.
The panels clearly show the progenitor at $\phi_1 = 0$ deg as it dominates the plot, and we can discern the stream extending from the progenitor in both $\phi_1$ directions.
We want to keep these features in our stream selection.
The following straight lines are designed by eye to describe these proper motion tracks:

\begin{align}
    \mu_{\phi_1, 0}(\phi_1) &= -0.09\phi_1 - 5, \text{ and }\label{eq:pmtracks1} \\
    \mu_{\phi_2, 0}(\phi_1) &= 0.26\phi_1 + 0.25\label{eq:pmtracks2},
\end{align}

\noindent and they are subtracted from the respective proper motion directions. 
They are verified to describe the stream by comparing them to the integrated GC orbit like we did for the distance modulus track in Eq. \ref{eq:DMgradient}.
We compute the intrinsic dispersion of the stream's $\mu_x-\mu_{x,0}(\phi_1)$ by only considering stars with $|\phi_1| > 0.3$ deg, and by assuming that the likelihood $\mathcal{L}$ can be described as a mixture of two Gaussians:

\begin{equation}\label{eq:pmtrack_mixturemodel}
    \mathcal{L}(\theta) = 
    f_\mathrm{x} \cdot \mathcal{N}(\mu_\mathrm{x}, \sqrt{\varepsilon_\mathrm{x}^2 + \sigma_\mathrm{x}^2}) + 
    (1-f_\mathrm{x}) \cdot \mathcal{N}(\mu_\mathrm{b}, \sqrt{\varepsilon_\mathrm{x}^2 + \sigma_\mathrm{b}^2}),
\end{equation}

\noindent where $\theta = (f_\mathrm{x}, \mu_\mathrm{x}, \sigma_\mathrm{x}, \mu_\mathrm{b}, \sigma_\mathrm{b})$.
There, $f_\mathrm{x}$ is the fraction of stream stars, $\mu_\mathrm{x}$ the mean of the stream proper motion Gaussian (where $\mu_\mathrm{x} \approx 0$ due to the track subtraction), $\sigma_\mathrm{x}$ is the width of the stream Gaussian, and $\mu_\mathrm{b}$ and $\sigma_\mathrm{b}$ are the same corresponding parameters for the background Gaussian. 
$\varepsilon_\mathrm{x}$ are the proper motion errors. 
The stream dispersion $\sigma_\mathrm{x}$ is what we are interested in, and we get it by applying a selection in one proper motion direction when computing the dispersion of the other direction. 
We first make a tentative selection around the $\mu_{\phi_1}$ peak.
We then use maximum likelihood estimation (MLE) to make a fit of the resulting $\mu_{\phi_2}$ distribution, using the likelihood given by Eq. \ref{eq:pmtrack_mixturemodel}, via Nelder-Mead optimization \citep{Gao12_Nelder-Mead} for all likelihood parameters simultaneously.
This gives a tentative value for $\sigma_{\phi_2}$.
We apply a selection of two $\sigma_{\phi_2}$ around $\mu_{\phi_2}$, and then repeat the fit of Eq. \ref{eq:pmtrack_mixturemodel} but now in $\mu_{\phi_1}$, which gives us a first guess for $\sigma_{\phi_1}$.
This is repeated until convergence, where a width of $2\sigma_x$ from the other proper motion direction fit is used to select stars. 
The final intrinsic proper motion dispersions are $\sigma_{\phi_1} = 0.11$ \masyr\ and $\sigma_{\phi_2} = 0.20$ \masyr.
The median error on the $\mu_{\phi_1}$ distribution when the $2 \sigma_{\phi_2}$ selection is applied is 0.3 \masyr, and the median error on the $\mu_{\phi_2}$ distribution when the $2 \sigma_{\phi_1}$ selection is made is 0.27 \masyr.
We therefore choose to describe our proper motion selection windows as:

\begin{align}
    |\mu_{\phi_1} - \mu_{\phi_1,0}(\phi_1)| &< n \sigma_{\phi_1} + 0.30, \text{ and }\label{eq:pmsel1} \\
    |\mu_{\phi_2} - \mu_{\phi_2,0}(\phi_1)| &< n \sigma_{\phi_2} + 0.27\label{eq:pmsel2},
\end{align}

\noindent where the tracks $\mu_{\phi_1,0}(\phi_1)$ and $\mu_{\phi_2,0}(\phi_1)$ are given in Eqs. \ref{eq:pmtracks1} and \ref{eq:pmtracks2}.

In the bottom row of Fig. \ref{fig:pmtracks} we show the photometric sample with all previous astrometric and photometric selections applied, as well as two dispersions around $\mu_{\phi_2}$ (for $\mu_{\phi_1}$, bottom left) and around $\mu_{\phi_1}$ (for $\mu_{\phi_2}$, bottom right), such that $n=2$ in Eqs. \ref{eq:pmsel1} and \ref{eq:pmsel2}.
The selection windows are shown as dashed magenta lines.
The stream proper motion slopes are clearly revealed in one proper motion direction when a selection is made in the other direction.
We additionally see the progenitor at $\phi_1=0$ deg as the distribution of stars extending above and below the proper motion tracks.
These proper motion distributions appear wide for progenitor stars because they are more numerous than stars in the stream, and because both populations have Gaussian proper motion distributions, there are more stars in the progenitor that can populate the tails of this Gaussian. 
In our analysis of the stream in Sec. \ref{sec:stream_modelling}, we remove stars within $|\phi_1| < 0.3$ deg to avoid the progenitor.

\subsection{Metallicity and radial velocity track selections}\label{sec:streamsel_fehvrad}

\begin{figure}
 \includegraphics{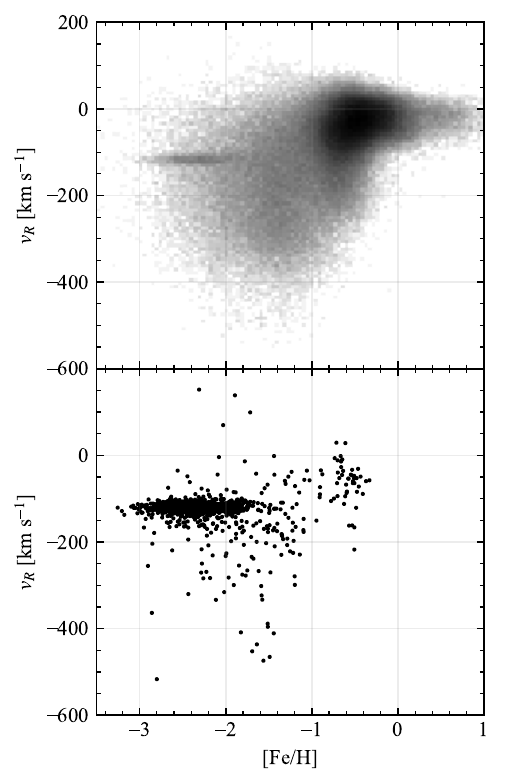}
 \caption{ 
          The metallicity and radial velocity distribution of the spectroscopic sample.  
          \textit{Top}: The entire sample, without any selections. 
          \textit{Bottom}: The sample, after photometric and astrometric selections have been applied, which makes the M92 overdensity around [Fe/H] $= -2.35$ dex and $v_R = -120$ \kms\ stand out even more strongly, showing the efficacy of these stream selections. 
          }
 \label{fig:specsample}
\end{figure}

Now that the photometric and astrometric stream selections have been finalised using the photometric sample, we can apply these selections to the DESI spectroscopic sample.
The radial velocity and metallicity distributions for the spectroscopic sample are shown in Fig. \ref{fig:specsample}, before and after these selections have been applied in the top and bottom panel respectively.
The top panel shows a strong peak from disc stars at ([Fe/H], $v_R$) $\approx (-0.7$ dex, 0 km s$^{-1}$) and a less prominent peak from M92 and its stream members at ([Fe/H], $v_R$) $\approx (-2.4$ dex, $-120$ km s$^{-1}$). 
This agrees with previous metallicity \citep{09Carretta_M92metallicity} and radial velocity \citep{Baumgardt19_GCorbitalparams} measurements for the progenitor.
In the bottom panel, we clearly see how the photometric and astrometric cuts isolate stream members and very efficiently removes contamination, with only some stars inside the M92 GC being removed because they fall outside the proper motion selection range (see the density of features around $\phi_1=0$ in the bottom panels of Fig. \ref{fig:pmtracks}).
However, some contamination remains, and we now introduce selections in metallicity and radial velocity to remove this contamination. 

Based on the bottom panel of Fig. \ref{fig:specsample}, we apply the following metallicity selection:

\begin{equation}\label{eq:metallicity_sel}
    -3.5 < \text{[Fe/H]} < -1.5.
\end{equation}

The metallicity dispersion of a GC is expected to be much smaller than the 2 dex range defined above. 
Earlier works found an upper limit of the metallicity scatter in GCs being less than 0.05 dex \citep{09Carretta_M92metallicity}. 
Later works have found a larger range in scatter, from 0.039 dex to 0.129 dex or up to 0.205 dex if $\omega$ Cen is considered \citep{Meszaros20_metallicitydispersion}, or even above 0.2 dex \citep{Geha26_GCDGs}. 
So one could be tempted to make a much thinner selection in metallicity than our 2 dex range.
However, in \citet[Table 7]{Koposov25_DESIDR1} the intrinsic metallicity scatter of M92 from the RVS pipeline, which is used in this work, was found to be 0.33 dex.
More importantly, most of the stars in our sample are faint and have large metallicity errors.
The large intrinsic metallicity scatter as well as large errors require a broader metallicity range in our selection to not wrongly remove stream stars.
Moreover, M92 shows complex abundances and metallicity distributions, indicating that it is not a normal GC but maybe the remnant of two merged GCs or the remnant of a dwarf galaxy nucleus (\citealt{Lee23_M92atypicalabundances, Lee24_M92helium}; Domínguez et al. in prep.), and a broader metallicity selection accounts for this. 

\begin{figure}
 \includegraphics{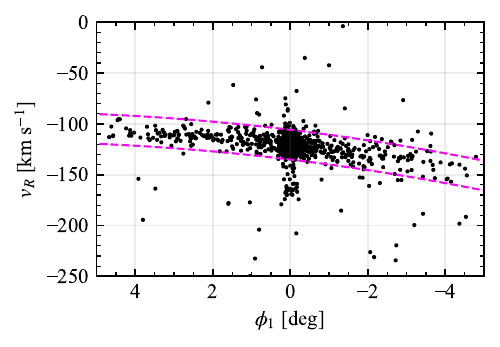}
 \caption{
          The radial velocity distribution as a function of stream longitude \lon\ of the spectroscopic sample after the astrometric selections in Eqs. \ref{eq:parallax_sel}, and \ref{eq:pmsel1} and \ref{eq:pmsel2} with $n=2$, as well as the metallicity selection given in Eq. \ref{eq:metallicity_sel}, has been applied.
          The magenta dashed lines show the velocity selection region, given by Eq. \ref{eq:vrsel} when $n=2$.
         }
 \label{fig:velsel}
 \end{figure}

After applying the selections in Eqs. \ref{eq:parallax_sel}, \ref{eq:pmsel1} and \ref{eq:pmsel2} with $n=2$, and \ref{eq:metallicity_sel} and the CMD selection, we get the radial velocity distribution as a function of stream longitude as shown in Fig. \ref{fig:velsel}.
It is interesting to note that the velocity dispersion seems larger for $\phi_1 < 0$ deg than for $\phi_1 > 0$ deg, which is also seen in the stream modelling in Figs \ref{fig:datafits} and \ref{fig:dispersions}.
The progenitor can additionally be seen as the stars extending above and below the velocity track, at $\phi_1=0$ deg.
A clear slope in radial velocity can be seen, which is approximated with the following track:

\begin{equation}\label{eq:vr_track}
    v_{R,0}(\phi_1) = -0.3 \phi_1^2 + 4.6 \phi_1 - 120.5.
\end{equation}

We apply a similar approach for the selection along the radial velocity track as we did for the proper motion selections described in Sec. \ref{sec:streamsel_pmtracks}.
By using the MLE approach with Nelder-Mead optimization, we find the intrinsic dispersion $\sigma_{v_R}$ of the distribution $v_R-v_{R,0}(\phi_1)$ by assuming the likelihood given in Eq. \ref{eq:pmtrack_mixturemodel}, when the previous selections are applied. 
We only consider the intrinsic dispersion of the stream by requiring $|\phi_1| < 0.3$ deg to remove bound stars.
This yields $\sigma_{v_R} = 5.14$ \kms.
It is larger than $\sigma_{\phi_1} = 4.33$ \kms\ but smaller than $\sigma_{\phi_2} = 7.87$ \kms, and we return to the differences in velocity dispersion in these three directions in Sec. \ref{sec:stream_modelling_results}.
Our $\sigma_{v_R}$ value is larger than those for many of GC streams \citep{Li22_S5onedozenstreams}, and the M92 stellar stream thus appears dynamically hot for having a GC progenitor. 
This could be a reflection of its orbital history, and possible be an effect from its pericentre being less than 1 kpc.
That larger velocity dispersions of GC streams could be caused by their orbital history has been discussed in the case of C-19 \citep{Yuan22_C19, Mohammed26_C19} and GD-1, which has a dynamically cold main stream and a hotter `cocoon' component (\citealt{Valluri25_GD1}, Jarvis et al. in prep.). 
We return to the orbital past of M92 as we create M92 mock streams in Sec. \ref{sec:mockstreams}. 

The selection region around the radial velocity track follows the same concept as for the proper motion selections, given in Eqs. \ref{eq:pmsel1} and \ref{eq:pmsel2}.
Applying the same selections to the data as described above, we find a mean radial velocity error of 4.36 \kms, and our radial velocity selection is thus defined as: 

\begin{equation}\label{eq:vrsel}
    |v_R - v_{R,0}(\phi_1)| < n \sigma_{v_R} + 4.36.
\end{equation}

\noindent The selection is shown in Fig. \ref{fig:velsel} as dashed magenta lines, when $n=2$.

\subsection{Stream morphology}\label{sec:streamsel_sky}

\begin{figure}
 \includegraphics{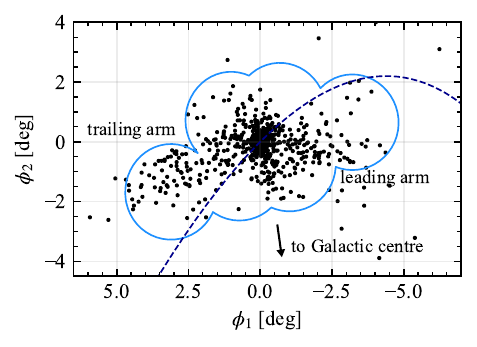}
 \caption{ 
          The stream morphology of the spectroscopic sample after selections in parallax, CMD, proper motions, metallicity and radial velocity have been applied.
          In blue, the outline of the DESI tertiary programme tiles are shown.
          The integrated GC orbit is shown as a dark blue dashed line.
          The direction of the Galactic centre has been marked with an arrow.
          }
 \label{fig:skysel_spec}
\end{figure}

To summarise, the final set of stream member selection criteria are as follows:
\begin{itemize}
    \item parallax (Eq. \ref{eq:parallax_sel}),
    \item CMD (Fig. \ref{fig:CMD_sel}),
    \item proper motions (Eqs. \ref{eq:pmsel1} and \ref{eq:pmsel2}),
    \item metallicity (Eq. \ref{eq:metallicity_sel}),
    \item and radial velocity (Eq. \ref{eq:vrsel}).
\end{itemize}
\noindent The projected sky morphology of the stream after applying the above selections to the spectroscopic sample is shown in Fig. \ref{fig:skysel_spec}.
The clear stream morphology as seen on the sky is a validation of our selection methods.
In the figure, we used $n=2$ in Eqs. \ref{eq:pmsel1}, \ref{eq:pmsel2} and \ref{eq:vrsel}.
These selections yield 1,188 stars, out of which 1,163 are within the six DESI tertiary programme tiles, which are outlined in blue (see also Fig. \ref{fig:phot_sample}).
These numbers include stars bound GC stars.

This is the first time M92 stream members have been characterised and selected using spectroscopy. 
Previous works by \citet{20Sollima_M92}, \citet{Thomas20_M92}, \citet{Ibata21_STREAMFINDER} and \citet{Chen25_StarStream} used \textit{Gaia} data and photometry to find stream members, though follow-up spectroscopy by \citet{Ibata21_STREAMFINDER} resulted in radial velocities for a few of their members. 
The extent of the stream on the sky is limited in our data because of the DESI tiles, but it is of note that the length we find roughly matches that by \citet{Thomas20_M92} who, unlike us, are not footprint restricted.
We observe a shorter stream than reported in \citet{Ibata21_STREAMFINDER}, who uses all-sky \textit{Gaia} data to find stream members.
We find that the stream is curved on the sky, and this curvature matches that found by \citet{Thomas20_M92} and \citet{Ibata21_STREAMFINDER}, though the stream found by the latter is tilted with respect to the stream found in this work.
Moreover, the stream in our data is not symmetric in \lon, as it is longer and broader at $\phi_1 > 0$ deg, which can partly be explained by the extent of the DESI tiles, though both \citet{Thomas20_M92} and \citet{Ibata21_STREAMFINDER} also find an asymmetric stream.
Overall, the morphology of the M92 stream in this work excellently matches that found by \citet{Chen25_StarStream}, who used all-sky \textit{Gaia} data: the thinness of the stream around $\phi_1 \sim -2.5$, the fluffiness at $\phi_1 > 0$ deg, the length, and the curvature are all in excellent agreement.

The stream can roughly be described by two parts: the more extended main part, and the two features above and below the progenitor at about $|\phi_1| = 0.5$ deg (though the feature below is more prominent).
In Fig. \ref{fig:skysel_spec}, the integrated GC orbit has been plotted as a dark blue dashed line, using the same method as for Fig. \ref{eq:DMgradient}. 
We can see that the main part of the stream is misaligned with its progenitor's orbit, which is expected for streams near their apocentres \citep{Bonaca25_streamsreview}.
M92 has just passed apocentre on its orbit around the MW.

The features above and below the progenitor constitute the so-called S-shape of the stream, which gets exaggerated at apocentre \citep{CapuzzoDolcetta05_streamSshape}.
The stream is moving from left to right in the figure and we have marked the direction of the Galactic centre with an arrow.
As expected, the S-shape is elongated along the direction of the Galactic centre, and the part directed towards the Galaxy (below the progenitor at $\phi_1<0$ deg) is bent towards the leading arm as it contains newly released stars that speed up ahead of the progenitor, while the part directed away from the Galaxy (above the progenitor at $\phi_1>0$ deg) is bent towards the trailing arm and contains slower moving stars \citep{Bonaca25_streamsreview}.
Stars decelerate at apocentre, and the inner stars that are elongated towards the MW are on tighter orbits and reach apocentre before the outer stars, meaning that they are more aligned with the Galactic direction because they have slowed down while the outer stars move faster \citep{Awad25_S5streams}, which could explain why the lower part of the S-shape is more pronounced than the upper.
Moreover, the S-shape in our data matches the features above and below the GC in N-body simulations by \citet{Thomas20_M92}.
The features we observe are also similar to those observed above and below NGC 1904, in which the S-shape is clearly visible due to our advantageous viewing angle of the stream: similarly to M92, NGC 1904 is on an eccentric orbit and close to apocentre \citep{Awad25_S5streams}.
Moreover, the S-shape feature extending below the progenitor is matched very well by the stream in \citet{Chen25_StarStream}.
We further discuss the differences between the M92 stream in this work and works found in the literature in Sec. \ref{sec:discussion_litcomparison_morphology}.

\section{Modelling the stream}\label{sec:stream_modelling}

The goal of this paper is to statistically compare the observed stream to mock streams in different barred potentials. 
We do this by computing summary statistics such as the means and standard deviations of the parameters stream latitude \lat, proper motions $\mu_{\phi_1}$ and $\mu_{\phi_2}$, and radial velocity $v_R$ in bins of \lon\ for both the observed and the mock streams.
In this section, we describe how this is done for the observed stream and the results.
We return to producing mock streams and computing summary statistics for them in Sec. \ref{sec:mockstreams}.
\subsection{The dataset}\label{sec:stream_modelling_datavec}

We start by applying the stream member selections to the spectroscopic sample using the same methods as developed in Sec. \ref{sec:streamid}.
To reliably model both the stream and the background stars, we now use a selection width of six standard deviations in proper motions and radial velocity ($n=6$ in Eqs. \ref{eq:pmsel1}, \ref{eq:pmsel2} and \ref{eq:vrsel}), as opposed to two standard deviations which was used in Fig. \ref{fig:skysel_spec}).
Stars outside the DESI tertiary programme tiles are removed to ensure even sky coverage, and to avoid the M92 progenitor, stars with $|\phi_1| < 0.3$ deg are also removed. 
In Appendix \ref{sec:appendixsdataforfits}, we show what the stream looks like in all four observables using these wider selections with the tile selection applied.
Because the observables have distributions that depend on \lon, we remove the tracks defined earlier. 
In proper motions and radial velocity, the tracks given in Eqs. \ref{eq:pmtracks1}, \ref{eq:pmtracks2} and \ref{eq:vr_track} are subtracted from \pmone, \pmtwo, and $v_R$ respectively.
From \lat, we remove the following stream track which has been designed by eye to match the stream:

\begin{equation}\label{eq:phi2gradient}
    \phi_{2,0}(\phi_1) = 0.021 \phi_1 - 0.076 \phi_1^2 - 0.01 \phi_1^3. 
\end{equation} 

\noindent The subtraction of the tracks ensure that our parameter distributions are not smeared and can accurately be described by Gaussian distributions.

The stream selections, tile footprint and \lon\ selections give us a vector of data $\vec{d} = [\phi_2, \mu_{\phi_1}, \mu_{\phi_2}, v_R]$ that have their respective tracks subtracted.
Now, each set of observables $\vec{d}_i$ is split into eight bins of \lon with steps of 1 deg, from $-4$ to 4 deg, creating $\vec{d}_{i,j}$, where $i$ corresponds to the observable and $j$ to the \lon\ bin.
Because the stream latitude track dips outside the DESI tiles in the bin $-4 < \phi_1 < -3$ deg, see Fig. \ref{fig:skysel_spec}, we exclude that bin for $\phi_2$. 
However there are still prominent peaks in proper motions and radial velocity in that bin, so it is kept for those observables.
This leaves us with 31 sets of $\vec{d}_{i,j}$, that have associated observational errors $\vec{\varepsilon}_{i,j}$ binned the same way. 
The likelihood and evaluation procedure are described in the next subsection.

\subsection{Mixture model and posterior sampling}

Each $\vec{d}_{i,j}$ can be described as a mixture model which contains a Gaussian stream distribution and a uniform background distribution.
Even though we go to lengths to remove contamination from the observations, we cannot assume that the DESI sample has perfect purity, which is why we need the latter component. 
The likelihood $\mathcal{L}$ we describe for $\vec{d}_{i,j}$ is thus:

\begin{equation}\label{eq:emceedatalikelihood}
    \mathcal{L}_{i,j} = f \cdot \mathcal{N}^*(\mu, \sqrt{\vec{\varepsilon}_{i,j}^2 + \sigma^2}) + (1-f) \cdot \mathcal{U},
\end{equation}

\noindent where $f$ the fraction of stream stars, $\mathcal{N}^*$ a truncated Gaussian, $\mu$ is the mean of the stream distribution and $\sigma$ the intrinsic dispersion, $\vec{\varepsilon}_{i,j}$ are the observational errors of each star corresponding to $\vec{d}_{i,j}$, and $\mathcal{U}$ is a uniform distribution over the entire parameter range.
The Gaussian distribution must be described as truncated, since we make selections in all observables $\vec{d}_i$.

All parameters of Eq. \ref{eq:emceedatalikelihood} have uniform priors, with the following ranges:
\begin{itemize}
\item $0 < f < 1$, 
\item $-3 < \mu_{\phi_2} < 3$ deg, 
\item $-5 < \mu_{\mu_{\phi_1}} < 5$ \masyr, 
\item $-8 < \mu_{\mu_{\phi_2}} < 8$ \masyr, 
\item $-400 < \mu_{v_R} < 400$ \kms,
\item $0.01< \sigma <$ the truncation window.
\end{itemize} 

The sampling of the posteriors is performed using \texttt{emcee} \citep{Foreman-Mackey13_emcee}.
We used 100 walkers and 10,000 steps out of which 5,000 were burn-in steps.
In each \lon\ bin $j$, we use the corresponding truncation edge for the observable $i$.
For the proper motions and radial velocity bins, the selection windows from Eqs. \ref{eq:pmsel1}, \ref{eq:pmsel2} and \ref{eq:vrsel} with $n=6$ mean that the same width is used for each $j$, which is 0.96 \masyr, 1.47 \masyr\ and 35.20 \kms\ respectively, and these are our truncation edges.
For \lat, we use the $\phi_1$-dependent edges of the DESI tiles which is different for each $j$ for the truncation edges.
From the sampled posterior, we compute the median and percentiles of the flat chain. 
This yields a list of 31 $\vec{\mu}_{i,j}$ and associated errors $\vec{\epsilon}_{i,j}$, which forms the summary statistic for the observed stream, and 31 mixing fractions $\vec{f}_{i,j}$ and dispersions $\vec{\sigma}_{i,j}$.
We discuss the resulting distributions of $\vec{\mu}$ and $\vec{\sigma}$ in the next subsection.

\subsection{Stream modelling results}\label{sec:stream_modelling_results}

\begin{figure*}
 \includegraphics{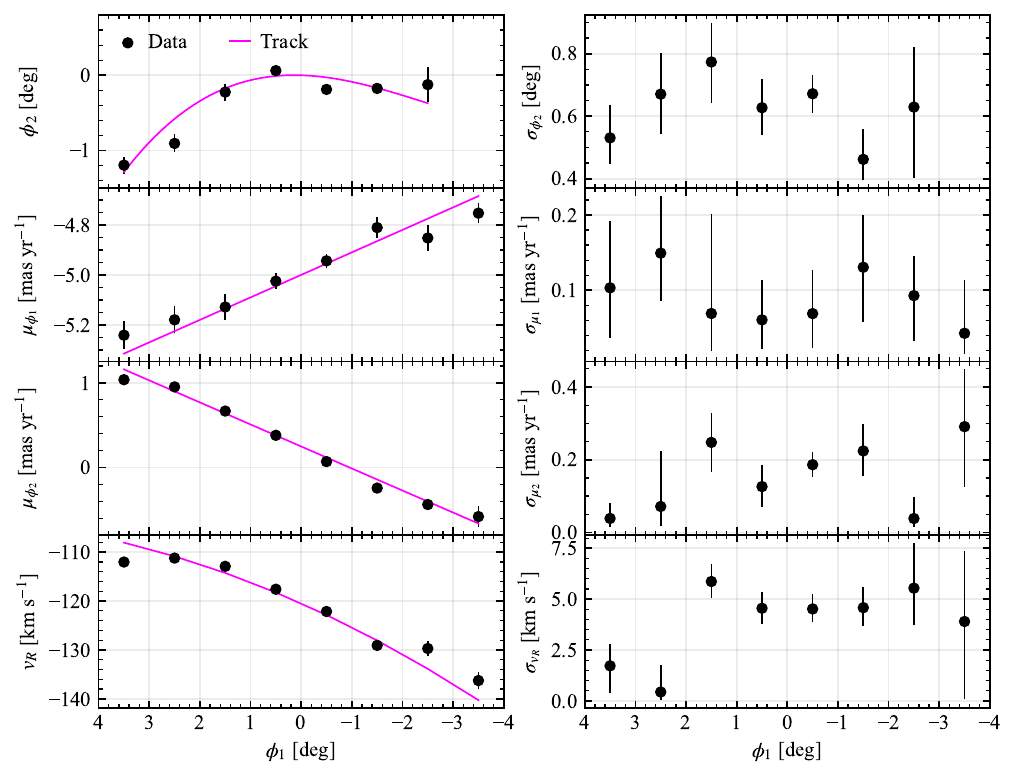}
 \caption{
          Means and dispersions of the four stream observables stream latitude (top row), proper motion along the stream longitude (middle top row), proper motion along the stream latitude (middle bottom row) and radial velocity (bottom row), obtained by fitting Eq. \ref{eq:emceedatalikelihood} to stream members in bins of \lon.
          Each row shows each of the four fitted observables as black points with corresponding error bars for each $\phi_1$ bin.
          Note that we exclude the $-4 < \phi_1 < -3$ deg bin for $\phi_2$ only.
          \textit{Left column}: The mean values, along with the tracks removed from the distributions given in magenta, which are Eqs. \ref{eq:phi2gradient}, \ref{eq:pmtracks1}, \ref{eq:pmtracks2} and \ref{eq:vr_track} for \lat, \pmone, \pmtwo\ and $v_R$ respectively.
          \textit{Right column}: The dispersion values.
         }
 \label{fig:datafits}
 \end{figure*}

The resulting $\vec{\mu}$ and $\vec{\sigma}$ from modelling the data as the mixture model given by Eq. \ref{eq:emceedatalikelihood}, are shown in Fig. \ref{fig:datafits}.
In the figure, each observable $i$ is presented in its own row, where the left column shows the means and the right column the dispersions.
In the left column, we also show the tracks removed from the distributions in magenta, which are Eqs. \ref{eq:phi2gradient}, \ref{eq:pmtracks1}, \ref{eq:pmtracks2} and \ref{eq:vr_track} for \lat, \pmone, \pmtwo\ and $v_R$ respectively.
The $\vec{\mu}_i$ agree very well with these tracks, which means that our selection of stream members around them is the correct method of identifying M92 stellar stream members.
We wish to point out the radial velocity dispersion distribution, which is constant around 5 \kms\ for the $\phi_1 < 2$ deg bins but decreases to less than half that value in the two leftmost bins, and this decrease in velocity dispersion can be identified by eye in Fig. \ref{fig:velsel}. 
These are the same \lon\ bins in which the radial velocity distribution plateaus and then increases.

\begin{figure}
 \includegraphics{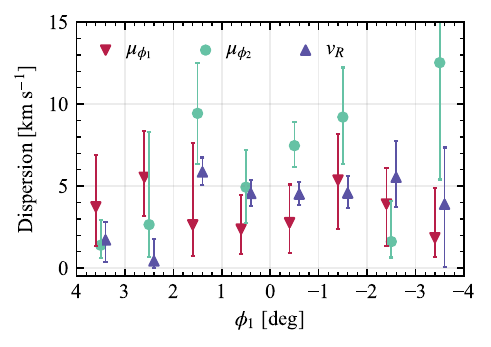}
 \caption{ 
          The velocity dispersions of the stream as seen in Fig. \ref{fig:datafits}, but with the proper motions converted into tangential velocity, assuming the stream distance modulus track given in Eq. \ref{eq:DMgradient}.
          For readability, $\mu_{\phi_1}$ and $v_R$ have been horizontally shifted to the left and the right, respectively.
          }
 \label{fig:dispersions}
\end{figure}

In Fig. \ref{fig:dispersions} we convert the proper motion dispersions into tangential velocity dispersions, assuming the stream distance modulus track given in Eq. \ref{eq:DMgradient}, and plot them together with the radial velocity dispersions.
For readability, $\mu_{\phi_1}$ and $v_R$ have been shifted horizontally by 0.1 deg.
Overall, the dispersions in the three directions agree within errors, with the exceptions of a few bins. 
This means that the stream has different kinematic behaviour in different directions, which could indicate a dramatic orbital history of the stream.
A similar mismatch in dispersion between the velocity directions has also been observed for the Orphan-Chenab stream, due to its past interactions with the Large Magellanic Cloud \citep{Koposov23_OrphanChenab}.

\section{Mock streams in barred potentials}\label{sec:mockstreams}

Now that we have identified M92 stellar stream members, we wish to compare the observed stream to mock streams generated in different barred MW potentials.

\subsection{Barred MW potentials}

For the orbit integration of our mock streams, we use the MW potential by \citet{24Hunter_barpotential}, following the implementation in the library for galaxy modelling \texttt{AGAMA}\footnote{\url{https://github.com/GalacticDynamics-Oxford/Agama/blob/master/py/example_mw_potential_hunter24.py}} \citet{Vasiliev19_AGAMA}.
The potential contains seven components in total: a central supermassive black hole, a nuclear star cluster, a nuclear stellar disc, a Galactic bar, a gas and a stellar disc, and a dark matter halo. 
The potential was designed to match the MW circular velocity curves by \citet{Eilers19_MWcircvelcurve} and \citet{Mroz19_MWcircvelcurve}, and observational constraints on the vertical acceleration at $|z| = 1.1$ kpc \citep{Bovy13_verticalacceleration} and at $|z| = 0.4$ kpc \citep{Widmark22_verticalacceleration}.
The potential's Galactic bar is based on an analytical model by \citet{22Sormani_MWbarmodel}, which is based on the model by \citet{Portail17_MWbarmodel}, which was constrained using the density of red clump giants and stellar kinematics in the bar and bulge region.
In the model, the bar consists of an X-shaped boxy bulge/bar component and a long bar component, with a total mass of $M_\mathrm{bar} = 1.83 \cdot 10^{10}$ M$_\odot$ \citep{22Sormani_MWbarmodel}.

All orbit integrations are performed in a right-handed Galactocentric Cartesian coordinate system, where the Sun sits on the negative $x$-axis.
When transforming these coordinates to equatorial coordinates we use built-in \texttt{AGAMA} routines.
The Sun is located at $(x, y, z) = (-8.122, 0, 0.0208)$ kpc and has corresponding velocities $(v_x, v_y, v_z) = (12.9, 245.6, 7.78)$ \kms\ \citep{Vasiliev19_AGAMA}.
To transform the coordinates into stream-centric coordinates, we use the M92 coordinate system defined in Sec. \ref{sec:data}.
In the Galactocentric Cartesian frame, the bar rotates clockwise.
This corresponds to a negative pattern speed with the \texttt{AGAMA} sign convention, but we present $|\Omega|$ in this paper to match the sign standard in the literature.

Because the Galactic bar is rotating, the bar angle with respect to the Galactic $x$-axis $\phi$ changes as a function of time $t$.
We define the dependence of the instantaneous $\phi$ on the present-day $\phi_0$ and bar pattern speed $\Omega$, and on the constant changing pattern speed $\dot \Omega$, as follows:

\begin{equation}\label{eq:thetaOmegaOmegadot}
    \phi(t) = \phi_0 + \Omega t + \frac{\dot \Omega t^2}{2},
\end{equation}

\noindent where $\phi_0=25$ deg, following the standard implementation in \texttt{AGAMA} but with a flipped sign to match the $\Omega$ sign convention. 
See e.g. the review by \citet{Hunt25_MWreview} for a discussion on the bar angle.
In this paper we consider the following ranges for the bar parameters: $0 < \Omega < 60$ \kmskpc\ and $-20 < \dot \Omega < 20$ \kmskpcGyr. 
A negative $\dot \Omega$ means the bar is decelerating and a positive $\dot \Omega$ means an accelerating bar.
This is a much larger range for both parameters than reasonable based on current constraints on both parameters, but we choose this range for completeness.
For the considered combinations of $\Omega$ and $\dot \Omega$ across the chosen parameter ranges, all other parameters of the potential such as e.g. mass distributions stay the same.
 
\subsection{Mock stream algorithm}

The mock streams are produced using the GC particle spray algorithm by \citet{Chen25_mockalgorithm}, which has been implemented in \texttt{AGAMA}\footnote{\url{https://github.com/ybillchen/particle_spray}}.
The \citet{Chen25_mockalgorithm} algorithm creates tracer particles outside the GC, thus avoiding the internal dynamics of the progenitor.
The tracer particles are placed around the progenitor's Lagrange points, with initial positions and velocities drawn from Gaussian distributions.
These distributions have been calibrated using N-body simulations of GC streams with varying progenitor masses, orbital radii, eccentricities and inclinations.

The algorithm is initiated by integrating the progenitor's orbit backward in the given potential from the present day, recording the progenitor's position at each time step. 
At each time step, a set of initial conditions for a stellar particle is created around each Lagrange point by drawing positions and velocities from the Gaussian distributions.
When the progenitor trajectory has been created, forward integration is done on the tracer particles by releasing two particles per time step as we move forward in time to the present day position, and at each subsequent timestep, the released particles feel the influence both of the MW and the progenitor's potential which we set to be a \citet{Plummer11_plummerpot} potential.
The mass loss rate of the progenitor is uniform in time and we assume that the progenitor's mass lost to the stream is negligible, so that its mass is constant throughout the orbit integrations.
This algorithm creates a stream at present day, with six-dimensional positions for each particle recorded. 

We initialise the progenitor's position and velocity at present day and define its Plummer potential using the M92 coordinates, mass and radius given in Table \ref{tab:progenitorinitialconds}. 
The stellar particles are integrated for 600 Myr, as the stream has a dynamical age of about 500 Myr \citep{Thomas20_M92}.
Moreover, stream particles integrated for a longer time than this would end up outside the DESI tiles, meaning that our footprint is limiting how far back in time we can observe the stream. 
In total 50,000 stellar particles are released from the progenitor.

\subsection{Comparing the observed stream to mock streams}

Now we turn to comparing the observed stream to mock streams, by first computing summary statistics for each mock stream. 
We create a vector $\vec{m}(\theta) = [\phi_2(\theta), \mu_{\phi_1}(\theta), \mu_{\phi_2}(\theta), v_R(\theta)]$, where $\theta = (\dot \Omega, \Omega)$ are the bar parameters for the considered potential.
Following the treatment of $\vec{d}_i$, from $\vec{m}(\theta)_i$ we subtract the tracks given in Eqs. \ref{eq:phi2gradient}, \ref{eq:pmtracks1}, \ref{eq:pmtracks2} and \ref{eq:vr_track} from $\phi_2(\theta)$, $\mu_{\phi_1}(\theta)$, $\mu_{\phi_2}(\theta)$ and $v_R(\theta)$ respectively.
We also remove particles outside the area covered by the DESI tertiary programme tiles and those that satisfy $|\phi_1| < 0.3$ deg.
No additional stream selections are applied, as all stellar particles in the mock stream are released from the progenitor GC.
We apply the same 31 bins in \lon\ to $\vec{m}(\theta)$ to create $\vec{m}(\theta)_{i,j}$.
The summary statistic we compute for $\vec{m}(\theta)_{i,j}$ is the mean of the distribution, $\vec{\mu}(\theta)_{i,j}$, by assuming a standard Gaussian.
We do not use the mixture model given in Eq. \ref{eq:emceedatalikelihood}, because we do not apply a stream selection on the mock streams which would otherwise require using a truncated Gaussian, and there is no contamination in the mocks that otherwise would require introducing a background component.

If our mock stream was perfect, we would expect that for the true bar parameters $\theta_\text{true}$, the data $\vec{\mu}_{i,j} \sim \mathcal{N}(\vec{\mu}_{i,j}(\theta_\text{true}), \vec{\varepsilon}_{i,j})$.
That would give us the likelihood function for the residual $\mathcal{L}(\theta) = \prod \mathcal{N}(\vec{\mu}_k - \vec{\mu}_k(\theta), \vec{\varepsilon}_k)$, where we sum over all 31 bins $k$.
The best bar parameters, $\theta$, are found by maximising $\mathcal{L}(\theta)$.
This can be done by taking the logarithm of $\mathcal{L}(\theta)$.
The logarithmic expression contains the term ln$(2 \pi \vec{\varepsilon}_k)$ that we can disregard as it is constant as a function of $\theta$. 
This leaves us with $-0.5\chi^2$.
We do not expect that our mock streams are perfect representations of the observed data, and indeed, the $\theta$ for which the reduced $\chi^2$ is the lowest, still returns $5.6$, which is a value far from one.
A more detailed discussion on the sources of these systematic differences are discussed in Sec. \ref{sec:discussion_comparison}.

To account for the $\chi^2 > 1$, we introduce a systematic error that depends on the given bar parameters and the specific observable considered, $\sigma_{\mathrm{sys},i}(\theta)$.
It is combined with the errors $\vec{\epsilon}$, which are the errors on the mean values in each \lon\ bin $\vec{\mu}$.
This gives us the following likelihood function defined per observable $i$:

\begin{equation*}
    \mathcal{L}_i(\sigma_{\mathrm{sys},i}(\theta)) = \prod_j \mathcal{N}(\vec{\mu}_j - \vec{\mu}_j(\theta), \sqrt{\vec{\epsilon}_j^2 + \sigma^2_{\mathrm{sys},i}(\theta)})
\end{equation*}

\noindent where $j$ sums to 7 for \lat\ and to 8 for the proper motions and radial velocities.

The likelihood in the above equation is evaluated using the MLE method via Nelder-Mead optimization to find $\sigma_{\mathrm{sys},i}(\theta)$ per $i$, and the resulting likelihoods are combined for a total likelihood across all observables.
This is done for all mock streams across the grid of bar parameters.
For each $\theta$, this thus yields in total four systematic errors, and we show the systematic errors as a function of bar parameters $\theta$ in Fig. \ref{sec:appendixsigmasys}, and a discussion of these values can be found in the accompanying Appendix \ref{sec:appendixsigmasys}.
We then find the systematic error vector $\vec{\sigma}_{\mathrm{sys,best}}(\theta)$ for which the likelihood is largest out of all mock streams, and keep the systematic errors fixed to those values for all $\theta$.
The bar parameter combination with the largest likelihood is $(\dot \Omega, \Omega) = (-0.125$ \kmskpcGyr, $28.875$ \kmskpc), and it gives us the vector of systematic errors $\vec{\sigma}_{\mathrm{sys,best}} = $ [0.281 deg, 0.025 \masyr, 0.090 \masyr, 1.980 \kms].
This $\vec{\sigma}_{\mathrm{sys,best}}$ ensures that the mock stream with the highest likelihood has a reduced $\chi^2 \sim 1$.
Fixing the systematic error, the likelihood we use to describe the data is given by:

\begin{equation}\label{eq:likelihood_grid}
    \mathcal{L}_i(\theta) = \prod_j \mathcal{N}(\vec{\mu}_j - \vec{\mu}_j(\theta), \sqrt{\vec{\epsilon}_j^2 + \sigma_{\mathrm{sys,best},i}^2})
\end{equation}

\noindent where $\sigma_{\mathrm{sys,best},i}$ is the element in $\vec{\sigma}_{\mathrm{sys,best}}$ which corresponds to a given observable $i$.
To get the probability across all four observables simultaneously, we sum all log($\mathcal{L}_i(\theta)$).

\subsection{Estimating $\Omega$ and $\dot \Omega$}

\begin{figure}
 \includegraphics{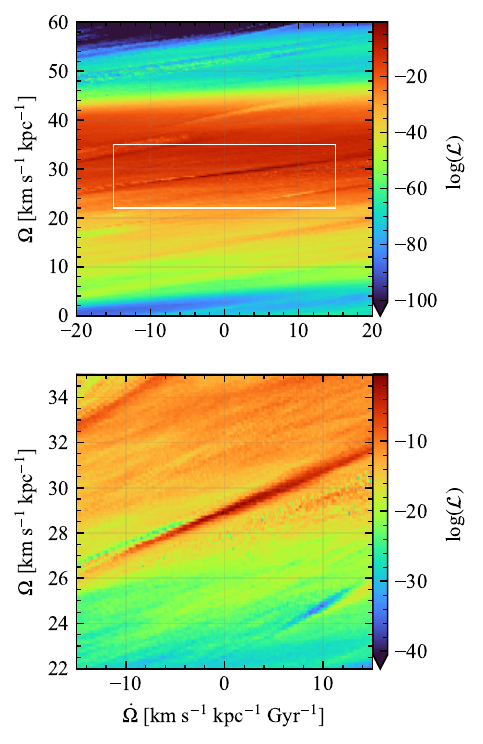}
 \caption{ 
          The log-likelihoods from Eq. \ref{eq:likelihood_grid} across the $(\dot \Omega, \Omega)$ grid of bar potential parameters in which we integrate our mock streams.
          Redder colours denote higher likelihood, and bluer colours denote lower likelihood.
          The bottom panel shows the area within the grey rectangle in the top panel (note also the change in colour scale).
          }
 \label{fig:logL_grid}
\end{figure}

The log-likelihood as a function of the bar parameters $\dot \Omega$ (on the x-axis) and $\Omega$ (on the y-axis) is shown for the entire parameter ranges in the upper panel of Fig. \ref{fig:logL_grid}. 
Red and blue denotes regions of high and low likelihood respectively.
In the bottom panel, we zoom in on the region shown by the white rectangle in the upper panel, as this region contains the highest likelihood feature which extends to the lower left and upper right of $(\dot \Omega, \Omega) = $ (0 \kmskpcGyr, 29 \kmskpc).
This ridge feature is a region of far greater likelihood of parameter space than other features across the grid.
Though it may appear as if e.g. the feature at about (--20 \kmskpcGyr, 32 \kmskpc) is equally likely, this is due to the logarithmic scaling of the likelihoods, as in reality these regions are many orders of magnitude less likely.
The ridge is thus the region of parameter space that best fits our M92 stream observation.

\begin{figure*}
 \includegraphics{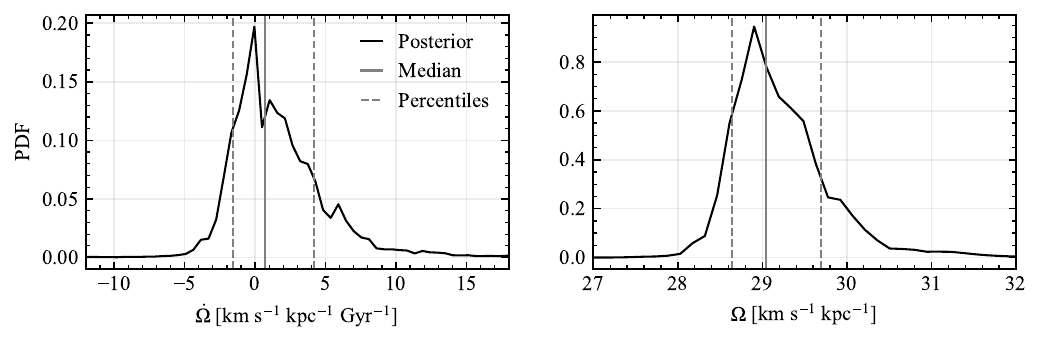}
 \caption{
          The posteriors on the bar parameters $\dot \Omega$ (\textit{left panel}) and $\Omega$ (\textit{right panel}), shown in solid black.
          The grey solid and dashed lines show the median and percentiles of each parameter respectively.
          }
 \label{fig:posteriors}
 \end{figure*}

To estimate the bar parameters $\dot \Omega$ and $\Omega$, we sample their joint posterior $p(\dot \Omega, \Omega | \vec{\mu}, \vec{\varepsilon}, \vec{\sigma}_\mathrm{sys,best})$ defined by the likelihood given in Eq. \ref{eq:likelihood_grid}, using the dynamic \citep{higson19_dynamicnestedsampling} nested sampler \citep{Skilling04_nestedsampling, skilling_nested_2006} \texttt{dynesty} \citep{Speagle20_dynesty, Koposov24_dynesty}, adopting the uniform priors $\dot \Omega \sim \mathcal{U}(-20, 20)$ and $\Omega \sim \mathcal{U}(0, 60)$.
We then estimate the 1-D marginal posterior for one bar parameter by marginalising over the other, using the uniform priors.
The estimated posteriors of the bar pattern speed and changing pattern speed are shown in Fig. \ref{fig:posteriors}.
The high-likelihood ridge described above can clearly be seen in the posterior of $\Omega$, and outside the panel ranges, the posteriors on both parameters are negligible.
The posteriors of both parameters are reasonably symmetric, so we compute their median with the errors taken as the 16th and 84th percentiles, which are shown in the panels as grey solid and grey dashed lines respectively.
These median and percentiles give us $\dot \Omega = 0.7^{+3.5}_{-2.3}$ \kmskpcGyr\ and $\Omega = 29.1^{+0.7}_{-0.4}$ \kmskpc, where the skewness on both parameters come from how extended and slanted the ridge is.
This result does not allow us to make a definitive statement on whether the MW bar is accelerating or decelerating.
In Appendix \ref{sec:appendixOmegaerrs} we compare the estimated errors on $\Omega$ and $\dot \Omega$, to the estimated errors when applying the above method to mock DESI streams in barred potentials of varying bar patterns speeds and accelerations.

\section{Discussion}\label{sec:discussion}

In this section, we first compare the observed stream with a mock stream generated in the potential with the best-fitting bar parameters in Sec. \ref{sec:discussion_comparison}, then in Sec. \ref{sec:discussion_logLpatterns} we discuss why the region of highly likely bar parameters in Fig. \ref{fig:logL_grid} (see especially the bottom panel) is favoured by the observed stream, and in Sec. \ref{sec:discussion_litcomparison} we compare our results to the literature.
 
\subsection{Mock stream generated in the best-fitting potential}\label{sec:discussion_comparison}

\begin{figure*}
 \includegraphics{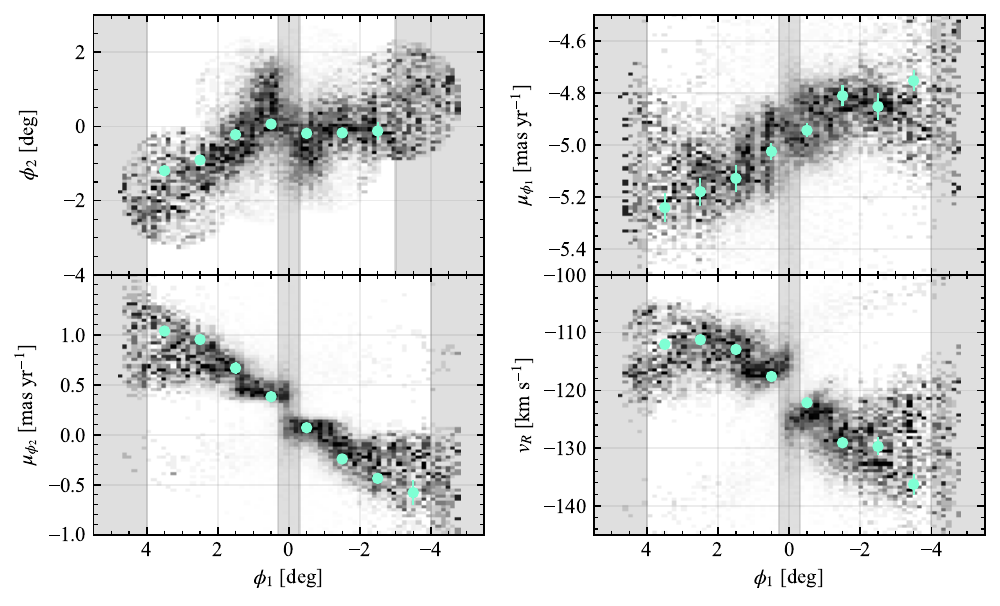}
 \caption{ 
          Mock stream generated in a potential with the best-fitting bar parameters, shown as a column-normalised histogram, 
          The results from the mixture modelling of the observed stream is shown as aquamarine points, which corresponds to the left column of Fig. \ref{fig:datafits}.
          Only mock stream particles within the DESI tiles have been selected.
          Regions of \lon\ that have not been used in the comparison of data and mock streams are greyed out, and these areas correspond to everything outside the bins in \lon\ as well as the central region of $|\phi_1| < 0.3$ deg.
         }
 \label{fig:bestmodel}
 \end{figure*}

When we evolve a progenitor in the barred potential with our estimated values of $\dot \Omega$ and $\Omega$, we get the mock stream presented in Fig. \ref{fig:bestmodel} after a DESI tertiary tile selection has been applied.
This figure shows stream latitude in the upper left panel, proper motion in stream longitude in the upper right panel, proper motion in stream latitude in the lower left panel, and radial velocity in the lower right panel.
The areas of \lon\ not used when comparing the observed stream to mock streams are greyed out at the left and right of each panel wherever $|\phi_1| > 4$ deg (except for \lat, where the $-4 < \phi_1 < -3$ bin is not used), and wherever $|\phi_1| < 0.3$ deg.
Points with error bars in aquamarine show the mixture model measurements from the observed stream, corresponding to the points in the left column of Fig. \ref{fig:datafits}.

We note three main differences between the mock stream and the observed stream.
The first difference is that in all three velocity directions, the slope is larger for the data than for the mock stream, and this is most easily seen for \pmtwo.
This is reflected in our vector of best-fitting systematic errors $\vec{\sigma}_{\mathrm{sys,best}}$, as our $\sigma_{\mathrm{sys,best},\mu_2} = 0.090$ \masyr\ $= 3.54$ \kms\ is larger than the other two velocity directions, i.e. $\sigma_{\mathrm{sys,best},\mu_1} = 0.025$ \masyr\ $= 0.98$ \kms\ and $\sigma_{\mathrm{sys,best},v_R} = 1.98$ \kms.
This systematic difference in slopes thus gets baked into the systematic errors that we include in our likelihood function for the residuals of the observed and mock streams in Eq. \ref{eq:likelihood_grid}.
It is also interesting to note that the comparison of dispersions of the velocity directions in Fig. \ref{fig:dispersions} show that along \pmtwo, the dispersion is on average the largest, and it is on average the smallest for the \pmone\ direction.
The large dispersion in \pmtwo\ as well as the larger mismatch between the mock stream and the observed stream in \pmtwo\, as indicated by $\sigma_{\mathrm{sys,best},\mu_2}$, compared to the other two velocity directions, points toward a past interaction with a gravitationally interacting object that is not included in our modelling, maybe similar to the interaction between the Orphan-Chenab stream and the Large Magellanic Cloud \citep{Koposov23_OrphanChenab}.

The second difference are the `ripples', i.e. the step-like variations around the main tracks along \lon\ in the velocity directions seen in the mock stream close to the progenitor location, which cannot be seen in the observed stream.
These are produced during the pericentric passages of the progenitor in its orbit around the MW centre, causing a change in dynamical heating during the orbit \citep{Chen25_mockalgorithm}.
The M92 GC completes five pericentric passages during our integration time of 600 Myr, and due to its ellipticity with a pericentre of $\sim0.3$ kpc and apocentre of $\sim10$ kpc, the difference in release velocities of the stars that become unbound at different points in the orbit is large, which makes these ripples more pronounced.
When errors randomly drawn from the distributions of the observed stream are added to the mock stream and it is downsampled to match the number of observed stream stars, these ripples are no longer perceptible. 

The mock stream exhibits what we refer to as `lobes', which are overdensities in \lat\ close to the progenitor location at $|\phi_1| = 0.5$ deg. 
These roughly match the increase of the mean stream distribution at $\phi_1 = 0.5$ and decrease at $\phi_1 = -0.5$ deg seen in Fig. \ref{fig:datafits}, and the feature at $\phi_1 = -0.5$ deg in Fig. \ref{fig:skysel_spec}, which is a stream feature often referred to as the S-shape \citep{CapuzzoDolcetta05_streamSshape}.
These features of the mock streams are similar to the distribution of stars close to the progenitor of the NGC 1904 stream, found in the Galactic centre-anticentre direction, which are due to the progenitor being close to apocentre on its highly eccentric orbit in combination with our viewing angle towards it \citep{Awad25_S5streams}.

The intensity of these lobes is the third difference, as this part of the stream is more pronounced in the mock stream than in the related S-shape of the observed stream.
These lobes are ubiquitous for all mock streams and more or less independent of the potential, as they consist of stars that were released during the most recent peri- and apocentres.
These may be exaggerated in our mock streams due to two factors. 
The first being the orbital eccentricity of M92, as \citet{Cook26_streammodelling} showed that the small pericentres of M92 may cause modelling errors on the sky plane when using particle spray algorithms to generate stream particles, instead of more realistic simulations such as N-body simulations.
The second factor may be the mass loss history of our mock stream.
The mock algorithm releases stellar particles at each equally spaced time step and does not alter its release conditions with distance from the host centre, which means that the lobes are populated to the same extent by stars from peri- and apocentre.
However, mass loss from the progenitor GC onto its stream is in reality larger at pericentre because the host potential is stronger at smaller Galactocentric distances \citep{Cai16_GCpericentremassloss, Ebrahimi19_GCpericentremassloss}. 
A more realistic mass loss history would then make these lobes less pronounced.
Because the lobes are primarily formed due to the mock stellar particles' release conditions and not the underlying potential in which they are integrated, they are ubiquitous for mock streams across the grid of bar parameters. 
The lobes thus do not cause our method to favour one region of bar parameter space.
Moreover, our modelling does not take the stellar surface density into account, meaning that the relatively large number of stars in the lobes do not affect the results when compared to the smaller amount of stars in the observed stream's S-shape.
and we do not believe that their existence biases our final results.

\subsection{The large probability region of bar parameter space}\label{sec:discussion_logLpatterns}

A striking feature of the log-likelihood as a function of bar parameters in Fig. \ref{fig:logL_grid} is the set of lines of high and low probability that appear across the plot. 
We zoom in on the most prominent of these lines in the bottom panel, extending from $(\dot \Omega, \Omega) =$ (0 \kmskpcGyr, $29$ \kmskpc) into the lower left and upper right sides of the plot.
This ridge in log$(\mathcal{L})$ is the source of the very peaked posteriors shown in Fig. \ref{fig:posteriors}.
We will now investigate why these combinations of bar parameters are so strongly preferred by the data and the properties shared by the potentials along this ridge.

\begin{figure*}
 \includegraphics{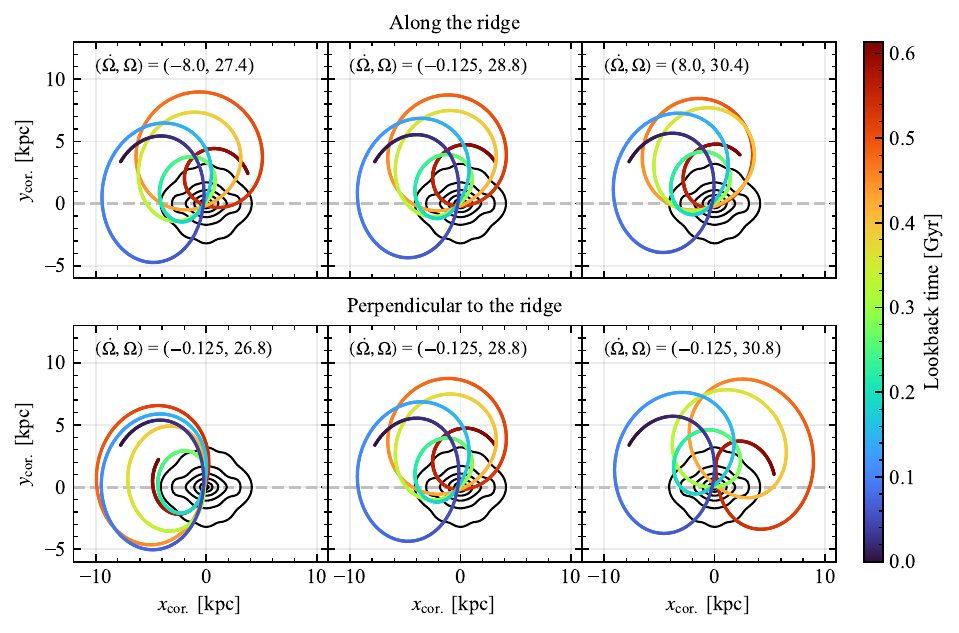}
 \caption{
          The stream progenitor orbit in different barred potentials, indicated at the top of each panel, in the bar co-rotating frame, colour-coded by lookback time.
          The bluest point shows the current position of M92.
          The bar density is shown as a black contour, and the bar's major axis is shown as a dashed grey line.
          \textit{Top row}: Orbits in potentials with bar parameters from the ridge of high likelihood bar parameters seen in Fig. \ref{fig:logL_grid}.
          \textit{Bottom row}: Orbits in potentials with a fixed bar deceleration but varying pattern speeds, above (left panel), on top of (middle panel), and below (right panel) the high-log$(\mathcal{L})$ ridge in Fig. \ref{fig:logL_grid}.
          The middle top and bottom panels are identical.
         }
 \label{fig:barorbit}
 \end{figure*}

In Fig. \ref{fig:barorbit} we compare orbits of M92 progenitors in different potentials in the bar co-rotating frame, with the bar density shown as black contours and the bar major axis along the x-axis marked with a dashed grey line.
The orbits are colour-coded by time, so that the most recent parts of the orbit are blue, and the furthest back in time are red.
The most recent parts of the orbits will be similar for different potentials, as we initialise all orbits at the current position of the progenitor from observations (see Table \ref{tab:progenitorinitialconds}).
The top row shows orbits in potentials with bar parameter combinations taken from the ridge feature in log$(\mathcal{L})$, and the bottom row shows orbits in potentials with a decelerating bar $\dot \Omega = -0.125$ \kmskpcGyr\ (which is where the $\dot \Omega$ posterior peaks), and values of $\Omega$ above, on top of, and below the ridge.

It is clear from Fig. \ref{fig:barorbit} that the orbits integrated in potentials with bar parameters along the ridge are almost identical (top row), but moving only slightly away from the ridge (bottom row) creates very different orbits.
The similarity between the orbits along the ridge explains why these bar parameters are preferred by the data, as the progenitor has undergone a similar orbital history in those potentials.
This was also pointed out by \citet{Yang25_Ophiuchusbar}, whose Fig. 12 shows a quantitative comparison of their observations of Ophiuchus to mock streams integrated in potentials with different combinations of $(\dot \Omega, \Omega)$.
This figure very similar to our Fig. \ref{fig:logL_grid}, and similarly shows large variations in how well the observed and the mock streams match, when the bar parameters change only slightly, something they also attribute to the bar's influence on the stream being so strong so that only a small change in its dynamics greatly affects the stream.

Moreover, orbits integrated in potentials with bar parameter combinations along the ridge all undergo their five pericentres at similar points relative to the bar.
For these orbits, the fifth and fourth pericentres happen on the bar major axis, which is not the case for orbits off the ridge.
This is also shown in the two leftmost panels in the top row of Fig. \ref{fig:appendix_deltatheta}, where the angular separation $\Delta \phi$ between the bar major axis and the progenitor at the fifth and fourth pericentres are plotted (see Appendix \ref{sec:appendix_deltatheta} for a more detailed description on this figure): orbits generated in potentials along the ridge, sketched with a dashed white line, have their two oldest pericentres aligned with the bar.
Comparing the slope of the ridge to the regions of $\Delta \phi=0$ or $180$ deg, we clearly see alignment, however considering a potential perpendicular to the ridge, the pericentres are no longer aligned with the bar.
This can also be seen in the left and right panels in the bottom row of Fig. \ref{fig:barorbit}, where the earliest pericentre occurs almost orthogonally to the bar.
The position of the pericentres with respect to the bar are important, as the stars experience the strongest influence from the host potential at pericentre \citep{Cai16_GCpericentremassloss, Ebrahimi19_GCpericentremassloss}, and if the orbits are aligned with the bar during these passages, its dynamical influence will be stronger.
In the top row of Fig. \ref{fig:barorbit}, this means that because the progenitor's orbits are aligned with the bar before and after the fourth and fifth pericentric passages, the GC feels a stronger effect from the bar than in the left and right panels of the bottom row, where these orbits are almost perpendicular to the bar.

\subsection{Comparison to the literature}\label{sec:discussion_litcomparison}

\subsubsection{M92 stream morphology}\label{sec:discussion_litcomparison_morphology}

In Sec. \ref{sec:streamsel_sky}, we compare the final stream sky distribution in our work and other works found in the literature, and we will now comment on likely causes for these differences.
The stream in the DESI data is very similar to that found by \citet{Thomas20_M92} who identified the stream using a matched-filter method and to that presented in \citet{Chen25_StarStream} where the \texttt{StarStream} algorithm was used. 
However, compared to the M92 stream found by \citet{Ibata21_STREAMFINDER}, we find a stream with a different orientation on the sky.
Because of this, out of their 32 M92 stream stars that are within the DESI tiles there are only 12 in common with our sample, and most of these are in the trailing tail ($\phi_1 > 0$ deg). 

\citet{Ibata21_STREAMFINDER} identified stellar members using the algorithm \texttt{STREAMFINDER}, which is designed to find stellar streams in \textit{Gaia} data. 
It does so by integrating the orbits of all stars in \textit{Gaia} and links them to stars on similar orbits in a potential containing a three-component disc, a bulge and a halo \citep{Dehnen98_MWmodel}. 
Where distance and velocity data for a given star are missing, a stellar population model is assumed and radial velocities are sampled from a range of values respectively \citep{Malhan18_STREAMFINDER}. 
The potential they use does not contain a bar, which is important because M92 occupies a region in energy--angular momentum space that experiences dynamical smearing by the MW bar \citep{Dillamore25_barsmearsELz} because its orbit is prograde \citep{Bajkova22_M92orbitalparams}, and assuming a bar-less potential for the orbit integration introduces the risk of incorrectly selecting stream members.
\citet{Thomas20_M92} argued that M92 could be affected by the MW bar based on the progenitor's orbit, and this is supported by the work in this paper. 
Therefore, the lack of a bar in the potential used in \texttt{STREAMFINDER} may explain why we see a different stream distribution compared to \citet{Ibata21_STREAMFINDER}.
In general, we believe that when characterising streams orbiting the MW with Galactocentric distances inside the region of significant dynamical influence from the bar \citep{Khoperskov20_barinfluence, Horta25_barspeed, Khalil25_barinfluence}, it is important to use a barred potential.

\subsubsection{The Galactic bar pattern speed and possible acceleration}

In this work, we measure the bar pattern speed and acceleration using a novel method where we compare an observed stream to mock streams in potentials with varying bar parameters, and we will now discuss how our best-fitting $\Omega$ differs from those found in the literature.
For a complete discussion on previous measurements of $\Omega$ and the associated methods, see \citet{Hunt25_MWreview}, and especially their Fig. 10 which presents a compilation of bar pattern speed measurements from the literature, showing that most works measure $\Omega \sim 35-40$ \kmskpc.
In short, previous works mainly depend on directly measuring the dynamics of stars in the inner MW.
If we compare our $\Omega=29.1$ \kmskpc\ to this compilation of $\Omega$ values, it is clear that we measure a slower bar pattern speed than most previous works.
The only exception is \citet{Horta25_barpatternspeed} who measure $\Omega = 24$ \kmskpc, and they suggest that their low pattern speed could be due to their method of disentangling the bar and the disc.

Our estimated $\dot \Omega$ does not allow us to make a definitive statement on whether or not the bar is accelerating or decelerating, since the error bars make our result consistent with both scenarios.
This large uncertainty likely come from how extended and linear the ridge feature is, which is why the posterior of $\Omega$ has the same shape as that of $\dot \Omega$.
The linearity of the ridge could stem from the linear dependence of $\Omega$ on $\dot \Omega$, as found when taking the time derivative of Eq. \ref{eq:thetaOmegaOmegadot}, for a fixed time $t$.
Considering the short orbit integration timescale in this work, fixing $t$ for our mock streams can be a valid approximation to explain the ridge's linearity.
 
That we do not conclusively measure a negative $\dot \Omega$ is in contrast with previous works in the literature that measure deceleration or requires a decelerating bar to explain a spectrum of observations of the MW.
\citet{Chiba21_bardeceleration} found that the pattern speed decelerates with $-4.5$ \kmskpcGyr.
An accelerating bar is in contrast with previous works in the literature that measure deceleration or requires a decelerating bar to explain a spectrum of observations of the MW.
\citet{Chiba21_bardeceleration} found that the pattern speed decelerates with $-4.5$ \kmskpcGyr.
Additionally, \citet{Yang25_Ophiuchusbar} required a decelerating bar to explain the Ophiuchus stream's morphology, \citet{Zhang25_bardispersedmetallicity} could explain the two distinct MW disc age-metallicity sequences with resonant dragging by the corotation of a decelerating bar, and \citet{Dillamore25_barspeed} also needed a decelerating bar to reproduce substructure in local MW halo radial velocity.
Comparing our work with that of \citet{Yang25_Ophiuchusbar}, we see that they successfully reproduce the Ophiuchus stream and especially its spur feature given a barred potential, instead of using the stream to probabilistically constrain the bar as in this work.
They integrate the progenitor for 3 Gyr, which is a much longer timescale than the 600 Myr in this work.
The works by \citet{Chiba21_bardeceleration}, \citet{Zhang25_bardispersedmetallicity} and \citet{Dillamore25_barspeed} use test particle simulations to compare against observed data, and they integrate these simulations for 9, 11.5 and 8 Gyr respectively, which is significantly longer than our considered timescale.
The timescale is important: though the bar should slow down due to angular momentum transfer to the disc and halo \citep{Debattista98_barsloseAM, Debattista00_barsloseAM, Athanassoula03_bardeceleration, MartinezValpuesta06_bardeceleration}, it was shown in \citet{Lokas14_baracceleration} that while this deceleration happens on long timescales, the bar can accelerate on shorter timescales due to e.g. tidal interactions, or by accretion events too \citep{Ash24_varyingbar}.
The Large Magellanic Cloud passed pericentre on its orbit around the MW only a few Myr ago \citep{Vasiliev24_LMCorbit} and it is possible that this tidal interaction has caused the MW's bar to suddenly accelerate.
Our shorter considered timescale could thus be why we measure $\dot \Omega$ in agreement with acceleration. 

Complications from directly measuring the kinematics of stars in the bar is not an issue in our work, as we measure how the bar affects the stream.
Instead, the biggest limitation in our work is the particle spray algorithm by \citet{Chen25_mockalgorithm}, which is tuned to the distributions of released stars from N-body simulations of GCs on different orbits around the MW.
These are evolved in a MW potential that contains a bulge, a disc and a dark matter halo, but no bar, which may impact our work as we assume a barred potential.
Moreover, the M92 GC has a more eccentric orbit \citep{Bajkova22_M92orbitalparams} than the range of GC orbit eccentricities in their simulations. 
It was shown in \citet{Cook26_streammodelling} that the small perigalacticon of M92 may cause errors in the generated stream particles when particle spray algorithms are used instead of more realistic stream simulations.
M92 additionally has a smaller apocentre \citep{Bajkova22_M92orbitalparams} than the range \citet{Chen25_mockalgorithm} use to tune their algorithm.
Given its eccentric orbit around the MW, M92 will experience a large increase in mass loss rate as it moves from apocentre to pericentre and this increase may be underestimated if its stream population has been tuned to more circular orbits \citep{Cai16_GCpericentremassloss, Ebrahimi19_GCpericentremassloss}.
Its eccentricity, combined with its large orbital inclination, may also mean that it experiences disc shocking \citep{Dehnen04_discshocking, Webb14_discshocking}, and the frequency of these may increase given its smaller apocentre.
Though we in Sec. \ref{sec:discussion_comparison} discuss the most apparent mismatches between our mock streams and the observed DESI stream that possibly arise due to the particle spray algorithm used, and conclude that they should not bias our final result, we cannot guarantee that using a more realistic stream simulation to generate our mock stream particles would change the estimated bar parameters in this work.

On top of the limitations from the particle spray algorithm used in this work, another limitation of our work is that we do not allow for the bar to change as it orbits the MW, as we use the potential in \citet{24Hunter_barpotential} which is based on the present-day MW and only let the bar rotate with time.
The bar length and strength can change as a function of time \citep{Bournaud02_varyingbar, Lokas14_baracceleration, Semczuk24_varyingbar}.
This can even occur on 100 Myr timescales, i.e. well within our stream integration time, when the bar aligns with the disc spiral arms \citep{Hilmi20_barparamfluctuations}.
In this work, we have ignored spiral arms, and this effect may be baked in to our current results.
A more realistic bar and the inclusion of spiral arms would potentially yield a slightly different $\Omega$ and $\dot \Omega$. 
Ideally, the bar length and strength, as well as the current bar angle $\phi_0$ that we have fixed as 25 deg but that has a range of values in the literature \citep{Hunt25_MWreview}, would be free parameters in our potential that are also constrained using the stream.
However, this likely requires a stream with a longer sky extent which is not limited by the observational footprint as in this work.

\section{Summary}\label{sec:conclusions}

In this work, we have used dedicated DESI observations centred on the M92 GC and the tidal stellar stream around it to identify stream members using photometry, astrometry and spectroscopy.
This is the first time spectroscopy has been used to characterise the M92 stellar stream.
We find a stream that has gradients in distance, proper motions and radial velocity as a function of stream longitude, all confirming the stream's existence. 

The stream is modelled in stream longitude bins assuming a mixture model, and compared to mock streams that have been integrated in barred MW potentials of varying bar pattern speeds $\Omega$ and changing pattern speeds $\dot \Omega$. 
This reveals a region of high likelihood in bar parameter space.
The posteriors of are estimated and we find $\Omega = 29.1^{+0.7}_{-0.4}$ \kmskpc\ and $\dot \Omega = 0.7^{+3.5}_{-2.3}$ \kmskpcGyr.
This is the first time a stellar stream has been used to probabilistically estimate parameters of the MW bar.

These parameters are tightly constrained because the orbit of the progenitor is very sensitive to the potential, as its trajectory takes it extremely close to the MW centre, and a small change in bar parameters can change the orbit with respect to the bar severely. 
Moreover, our results are consistent both with an accelerating and a decelerating bar. 
The fact that we do not decisively measure deceleration like in previous works could be due to the recent pericentric passage of the Large Magellanic Cloud, as cosmological simulations have shown that tidal interactions can cause a bar to accelerate \citep{Lokas14_baracceleration}.

Future studies of the M92 stellar stream would benefit from a larger observational region than the six DESI tiles used in this work, as it would allow for dynamically older parts of the stream to be characterised.
These dynamically older parts will have experienced the kinematic effect of the rotating bar during a longer timescale, and could thus help in constraining the past rotation of the bar, which could improve the bar constraints.
Additionally including a bar whose length and strength varies with time would strengthen these kinematic constraints.
Future work would also benefit greatly from generating M92 particles using N-body simulations instead of particle spray algorithms, as the latter can be limiting \citep{Fardal15_mockstreams, Chen25_mockalgorithm}, especially for streams with progenitor orbits that have small pericentres and thus experience strong tidal forces from the host potential such as M92 \citep{Cook26_streammodelling}.
Additionally, applying this paper's methodology to a homogeneously selected set of stellar streams with spectroscopic information would allow us to understand the effect of the bar on streams as a function of stream orbital parameters, and with a larger dataset, we could get the opportunity to constrain other bar parameters than considered in this work such as its length and angle.
To summarise, the method developed in this work opens up a new avenue of constraining the Galactic bar and its interactions with stellar streams.

\section*{Acknowledgements}

SK acknowledges support from the Science \& Technology Facilities Council (STFC) grant ST/Y001001/1.
GT acknowledges support from the Agencia Estatal de Investigaci\'on del Ministerio de Ciencia en Innovaci\'on (AEI-MICIN) under grant number PID2020-118778GB-I00/10.13039/501100011033 and the grant RYC2024-051016-I funded by MCIN/AEI/10.13039/501100011033 and by the European Social Fund Plus.

This work made use of the Python packages \texttt{numpy} \citep{harris20_numpy}, \texttt{matplotlib} \citep{Hunter07_matplotlib}, \texttt{scipy} \citep{Virtanen20_scipy}, \texttt{gala} \citep{Price-Whelan17_gala, adrian_price_whelan_2020_4159870}, and \texttt{astropy} \citep{Astropy13_astropy, Astropy18_astropy, Astropy22_astropy}. 
This paper made use of the Whole Sky Database (WSDB) created and maintained by Sergey Koposov at the Institute of Astronomy, Cambridge with financial support from the Science \& Technology Facilities Council (STFC) and the European Research Council (ERC).

This material is based upon work supported by the U.S. Department of Energy (DOE), Office of Science, Office of High-Energy Physics, under Contract No. DE–AC02–05CH11231, and by the National Energy Research Scientific Computing Center, a DOE Office of Science User Facility under the same contract. Additional support for DESI was provided by the U.S. National Science Foundation (NSF), Division of Astronomical Sciences under Contract No. AST-0950945 to the NSF’s National Optical-Infrared Astronomy Research Laboratory; the Science and Technology Facilities Council of the United Kingdom; the Gordon and Betty Moore Foundation; the Heising-Simons Foundation; the French Alternative Energies and Atomic Energy Commission (CEA); the Secretariat of Science, Humanities, Technology and Innovation (SECIHTI) of Mexico; the Ministry of Science, Innovation and Universities of Spain (MICIU/AEI/10.13039/501100011033), and by the DESI Member Institutions: \url{https://www.desi.lbl.gov/collaborating-institutions}. Any opinions, findings, and conclusions or recommendations expressed in this material are those of the author(s) and do not necessarily reflect the views of the U. S. National Science Foundation, the U. S. Department of Energy, or any of the listed funding agencies.

The authors are honored to be permitted to conduct scientific research on I'oligam Du'ag (Kitt Peak), a mountain with particular significance to the Tohono O’odham Nation.

\section*{Data Availability}

Upon acceptance of this paper, we will make the following available: the objects observed as part of the dedicated M92 DESI tertiary programme, a tutorial on how to select stream members according to the methods in this paper, the stream modelling results presented in Fig. \ref{fig:datafits}, the data behind all figures in this paper and a tutorial on how to recreate these figures.


\bibliographystyle{mnras}
\bibliography{bib}



\appendix

\section{Stream selection for modelling}\label{sec:appendixsdataforfits}

\begin{figure*}
 \includegraphics{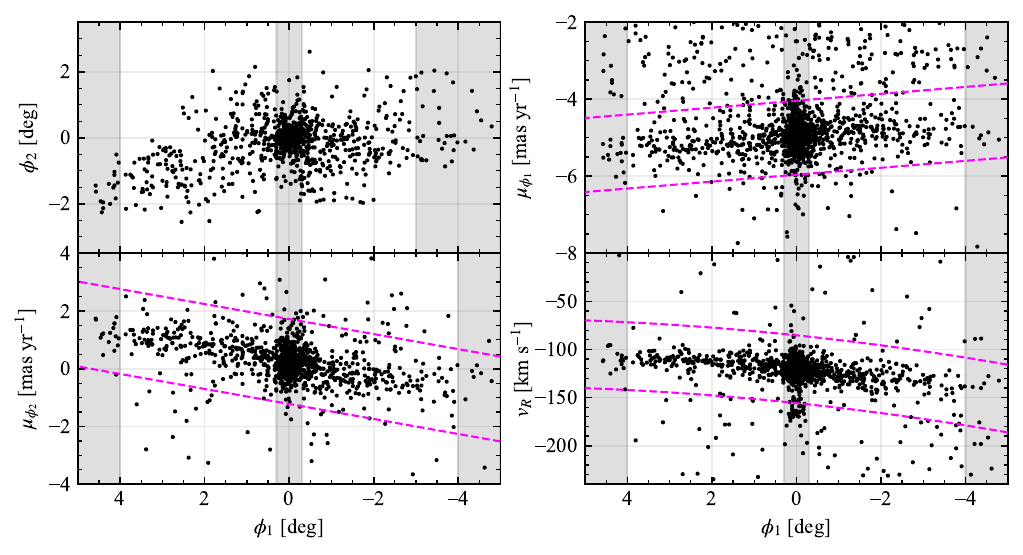}
 \caption{
          The data used for evaluating Eq. \ref{eq:emceedatalikelihood} in all four observables, by setting $n = 6$ in Eqs. \ref{eq:pmsel1}, \ref{eq:pmsel2} and \ref{eq:vrsel}.
          The dashed magenta lines show the selection region for that specific observable, with each panel having the selections from only the other panels applied.
          The greyed regions show the regions in $\phi_1$ not used in the likelihood evaluation.
          \textit{Top left}: $\phi_2$.
          \textit{Top right}: $\mu_{\phi_1}$. 
          \textit{Bottom left}: $\mu_{\phi_2}$. 
          \textit{Bottom right}: $v_R$. 
        }
 \label{fig:appendix_dataforfits}
 \end{figure*}

When we evaluate Eq. \ref{eq:emceedatalikelihood} on the observed stream, we use a stream selection with a wide cut around the proper motions and radial velocity, to ensure that enough stars are included in the sample for a good assessment of the background component.
Specifically, we set $n=6$ in Eqs. \ref{eq:pmsel1}, \ref{eq:pmsel2} and \ref{eq:vrsel} (as opposed to $n=2$ seen in Figs. \ref{fig:pmtracks} and \ref{fig:velsel}).
The data in all four observables using this selection is shown in Fig. \ref{fig:appendix_dataforfits}, with the regions not included in our \lon\ bins greyed out, and the selection ranges in the velocity directions are shown as dashed magenta lines, and as shown by the upper left panel, only stars inside the DESI tiles are used.
In each panel, all selections from the other panels have been applied, but not that panel itself (with the exception of $\phi_2$ where the tiling selection has been applied).
This shows that even with this broad selection, a stream is clearly visible. 
The stars outside the greyed regions with the selections in all panels applied are used to create the data vector $\vec{d}$ described in Sec. \ref{sec:stream_modelling_datavec}.

\section{Systematic errors for the observables}\label{sec:appendixsigmasys}

\begin{figure*}
 \includegraphics{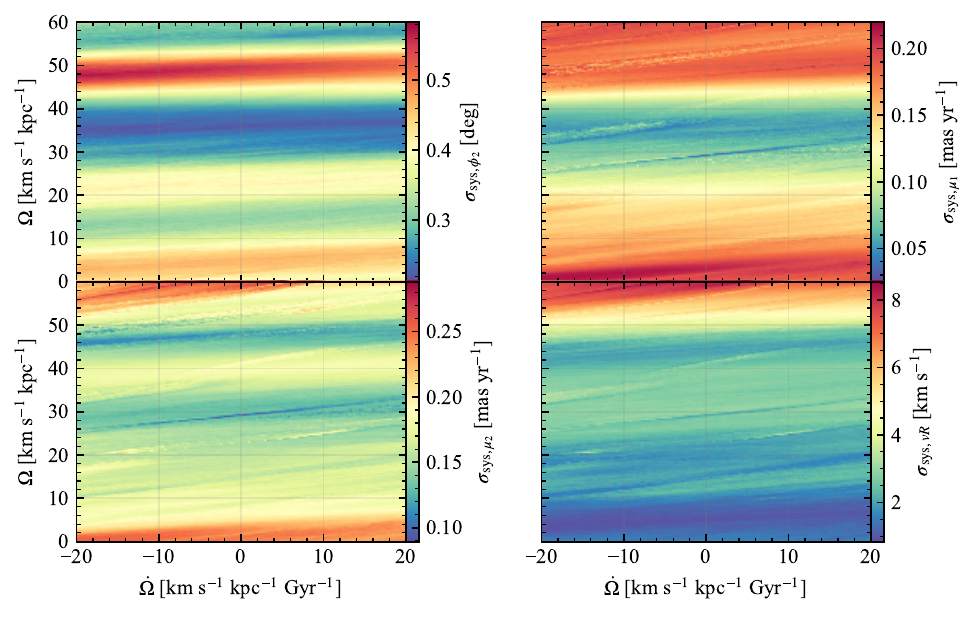}
 \caption{
          The $\sigma_\mathrm{sys}$ given by Eq. \ref{eq:likelihood_grid} presented per observable, as a function of bar parameters $\theta$.
          \textit{Top left}: The $\phi_2$ systematic error.
          \textit{Top right}: The $\mu_{\phi_1}$ systematic error. 
          \textit{Bottom left}: The $\mu_{\phi_2}$ systematic error. 
          \textit{Bottom right}: The $v_R$ systematic error. 
        }
 \label{fig:appendix_syssigma}
 \end{figure*}

Eq. \ref{eq:likelihood_grid} contains a systematic error $\sigma_{\text{sys},i}(\theta)$ defined per observable $i$. 
For every set of bar potential parameters $\theta$ we fit this variable, so that each mock stream has a set of $\vec{\sigma}_{\text{sys}}$ values associated with its underlying potential. 
The systematic errors are shown as a function of $\theta$, split into the four observables, in Fig. \ref{fig:appendix_syssigma}.
The three velocity observables show the same streak patterns as each other and as Fig. \ref{fig:logL_grid} whereas the $\phi_2$ systematic errors show a different pattern, with less clear streaks, and almost no dependence on $\dot \Omega$ and instead broad bands across as a function of $\theta$.
We believe that this difference in patterns to the most part can be explained by the more complex selection on $\phi_2$ than the velocity observables, as it will be affected by the DESI tertiary programme tiles.

\section{Constraints on $\Omega$ and $\dot \Omega$ from mock DESI streams}\label{sec:appendixOmegaerrs}

The thinness of the ridge in log$(\mathcal{L})$, seen in Fig. \ref{fig:logL_grid}, is translated into a thin posterior for the bar pattern speed, which is why the error on $\Omega$ is so small.
We now test if this error is realistic or underestimated, by applying our pipeline to mock DESI data.
This is done by generating 1,000 mock streams in potentials with random combinations of $\dot \Omega$ and $\Omega$ from the same ranges used in Fig. \ref{fig:logL_grid}.
DESI-like errors in proper motions and radial velocities are randomly drawn from the observed stream and applied to each final mock stream.
For every combination of bar parameters, this creates a mock DESI stream.
After applying errors, we only select mock DESI particles that are inside the DESI tiles and have $|\phi_1| > 0.3$ deg.
The number of generated mock stars is chosen so that there is roughly the same amount of final stellar particles as in the observed stream. 
The DESI mock streams are binned in the same way as described in Sec. \ref{sec:stream_modelling_datavec} to create a $\vec{d}$, and the summary statistic in each $\vec{d}_{i,j}$ is computed by assuming a standard Gaussian distribution. 
The mixture model used for the DESI data given in Eq. \ref{eq:emceedatalikelihood} is not used for the mock DESI streams, as we know that there will be no background component contaminating these streams, and the streams are not truncated because we do not apply cuts to the mock streams again because we know that there is no background contamination.

For each of the 1,000 mock DESI streams, we sampled the marginal posteriors of $\dot \Omega$ and $\Omega$ with the same method as we used for the observational data.
The median, i.e. the inferred value, of these posteriors are compared to the true values for a given potential, and we now consider the distributions of the residuals.
The median of $\Omega_\text{true}-\Omega_\text{inferred}$ is $-0.2$ with a standard deviation of 2.5 \kmskpc.
Since the median formal error on $\Omega$ from the data is 0.4 \kmskpc\ for the upper error and 0.7 \kmskpc\ for the lower error, we believe that 2.5 \kmskpc might be a more realistic error.
$\dot \Omega_\text{true}$ is not constrained for any mock stream because all the posteriors of $\dot \Omega_\text{inferred}$ are more or less flat.
We believe that it is more difficult to constrain the bar parameters using the mock DESI streams than the observed stream because the mock streams behave differently than the observed stream (which is what prompted us to introduce $\sigma_\mathrm{sys}$ in Eq. \ref{eq:likelihood_grid}), which comes from a combination of the mock stream algorithm and the evolution in the host potential.
For a discussion on the mock stream algorithm, see Sec. \ref{sec:discussion_comparison}.
Regarding the evolution in the host potential, the stream progenitor experiences pericentres as small as 0.3 kpc, and at these Galactocentric distances, the exact shape of the potential is crucial.
Even though the barred potential we use is carefully designed to contain all components of the inner MW \citep{24Hunter_barpotential}, there could be some details in the real Galaxy that are not perfectly described by the potential, which could show up as differences in the observed stream and the mock DESI streams. 

\section{Bar and progenitor alignments at peri- and apocentre}\label{sec:appendix_deltatheta}

\begin{figure*}
 \includegraphics{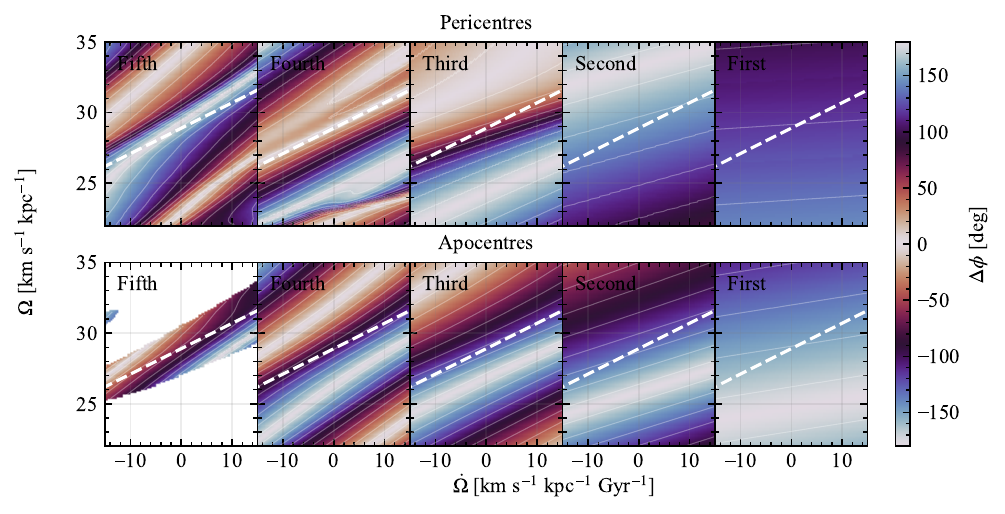}
 \caption{
          The angular separation $\Delta \phi$ between the bar major axis and the GC stream progenitor, in different barred potentials.
          The range in $(\dot \Omega, \Omega)$ is the same as in the bottom panel of Fig. \ref{eq:likelihood_grid}.
          The ridge feature of high likelihood bar parameters seen in that panel is here sketched as a dashed white line.
          Alignment occurs when $\Delta \phi = 0$ or 180 deg, as both sides of the bar are equally influential on the progenitor.
          Lookback time decreases going from left to right, and as the progenitor has just passed apocentre, the bottom right panel is the closest to present time.
          Between each column, there is about 100 Myr that have passed, as we integrate the progenitor for only 600 Myr in total.
          \textit{Top row:} The angular separations at each pericentre.
          \textit{Bottom row:} The angular separations at each apocentre.
          The bottom left panel has empty, white regions as the progenitor only undergoes four apocentric passages in some potentials.
        }
 \label{fig:appendix_deltatheta}
 \end{figure*}

The sharp ridge feature in the log-likelihood as a function of bar parameters (see Fig. \ref{fig:logL_grid}) which translates to the very peaked posteriors of the bar pattern speed (see the right panel of Fig. \ref{fig:posteriors}), can be explained by the stream progenitor experiencing similar orbital histories in these high-likelihood potentials (see Fig. \ref{fig:barorbit}).
The most important feature of the orbits in these potentials is that the progenitor has its earliest two pericentres aligned with the bar major axis.
In Fig. \ref{fig:barorbit}, this can be seen in the upper row where the red and orange parts of the orbit align with the bar major axis, compared to the lower row where potentials that have lower likelihoods create different progenitor and bar alignments during pericentre.

In Fig. \ref{fig:appendix_deltatheta} we show the distribution of angular differences $\Delta \phi$ between the bar major axis and the progenitor as a function of bar parameters, at each of the five pericentres the progenitor experiences in the top row and each of the five apocentres in the bottom row. 
The range in bar parameters is the same as the bottom panel of Fig. \ref{fig:logL_grid}.
The peri- and apocentres furthest back in time are to the left, and time progresses forward to the right (corresponding to the red to blue transition in Fig. \ref{fig:barorbit}).
Not all potentials produce five apocentres during our integration time of 600 Myr, which is why there is a white region in the bottom left panel; for these potentials, only four apocentres occur for the progenitor.
To guide the eye, the rough location of the high-likelihood ridge has been marked with a dashed white line and contours in $\Delta \phi$ is included.

Because the orbit integration backwards in time begins in all potentials from the same present-day position of the progenitor, orbits integrated in different potentials become more different the further back in time we go, and so the most structure in Fig. \ref{fig:appendix_deltatheta} can be seen for the earliest peri- and apocentres.
A clear feature for potentials along the ridge is that for during the fifth and fourth pericentric passages (leftmost and second leftmost top panels in Fig. \ref{fig:appendix_deltatheta}), alignments with the bar major axis occurs. 
Here we define alignment as when $\Delta \phi = 0$ deg or 180 deg, as both ends of the bar are considered equally influential on the orbit.
This is important, as pericentre is when the progenitor is closest to the Galactic centre, and if this occurs along the bar major axis, the progenitor has felt the influence of the bar for the longest possible time during its orbit.
These are thus the regions where the bar has the strongest influence on the progenitor.
A tiny change in the bar rotation parameters will shift the orbit ever so slightly away from this strong bar influence, which at these small Galactocentric distances causes a great change in the final orbit distribution, which affects the final stream.
The precise position of the progenitor with respect to the bar as it orbits the MW is thus encoded in the final stream, and allows us to separate which set of bar parameters are true for the MW. 

\vspace{1em}

\noindent $^{1}$ \textit{Institute for Astronomy, University of Edinburgh, Royal Observatory, Blackford Hill, Edinburgh EH9 3HJ, UK}\\
$^{2}$ \textit{Institute of Astronomy, University of Cambridge, Madingley Road, Cambridge CB3 0HA, UK}\\
$^{3}$ \textit{Department of Astronomy \& Astrophysics, University of Toronto, Toronto, ON M5S 3H4, Canada}\\
$^{4}$ \textit{Department of Astronomy \& Astrophysics, University of California, Santa Cruz, 1156 High Street, Santa Cruz, CA 95064, USA}\\
$^{5}$ \textit{NSF NOIRLab, 950 N. Cherry Ave., Tucson, AZ 85719, USA}\\
$^{6}$ \textit{Universidad de La Laguna, Av. Astrof\'{\i}sico Francisco S\'{a}nchez E-38205 La Laguna, Santa Cruz de Tenerife, Spain} \\
$^{7}$ \textit{Instituto de Astrof\'{\i}sica de Canarias, C/ V\'{\i}a L\'{a}ctea, s/n, E-38205 La Laguna, Tenerife, Spain} \\
$^{8}$ \textit{Department of Astronomy, University of Michigan, Ann Arbor, MI 48109, USA}\\
$^{9}$ \textit{Institute of Astronomy and Department of Physics, National Tsing Hua University, 101 Kuang-Fu Rd. Sec. 2, Hsinchu 30013, Taiwan}\\
$^{10}$ \textit{Center for Informatics and Computation in Astronomy, National Tsing Hua University, Hsinchu 30013, Taiwan}\\
$^{11}$ \textit{Lund Observatory, Division of Astrophysics, Department of Physics, Lund University, SE-221 00 Lund, Sweden}\\
$^{12}$ \textit{Observat\'{o}rio Nacional, Rio de Janeiro 20921-400, Brasil}\\
$^{13}$ \textit{Lawrence Berkeley National Laboratory, 1 Cyclotron Road, Berkeley, CA 94720, USA}\\
$^{14}$ \textit{Department of Physics, Boston University, 590 Commonwealth Avenue, Boston, MA 02215 USA}\\
$^{15}$ \textit{Dipartimento di Fisica ``Aldo Pontremoli'', Universit\`a degli Studi di Milano, Via Celoria 16, I-20133 Milano, Italy}\\
$^{16}$ \textit{INAF-Osservatorio Astronomico di Brera, Via Brera 28, 20122 Milano, Italy}\\
$^{17}$ \textit{Department of Physics \& Astronomy, University College London, Gower Street, London, WC1E 6BT, UK}\\
$^{18}$ \textit{Instituto de F\'{\i}sica, Universidad Nacional Aut\'{o}noma de M\'{e}xico,  Circuito de la Investigaci\'{o}n Cient\'{\i}fica, Ciudad Universitaria, Cd. de M\'{e}xico  C.~P.~04510,  M\'{e}xico}\\
$^{19}$ \textit{Departamento de F\'isica, Universidad de los Andes, Cra. 1 No. 18A-10, Edificio Ip, CP 111711, Bogot\'a, Colombia}\\
$^{20}$ \textit{Observatorio Astron\'omico, Universidad de los Andes, Cra. 1 No. 18A-10, Edificio H, CP 111711 Bogot\'a, Colombia}\\
$^{21}$ \textit{University of Virginia, Department of Astronomy, Charlottesville, VA 22904, USA}\\
$^{22}$ \textit{Fermi National Accelerator Laboratory, PO Box 500, Batavia, IL 60510, USA}\\
$^{23}$ \textit{Sorbonne Universit\'{e}, CNRS/IN2P3, Laboratoire de Physique Nucl\'{e}aire et de Hautes Energies (LPNHE), FR-75005 Paris, France}\\
$^{24}$ \textit{Instituci\'{o} Catalana de Recerca i Estudis Avan\c{c}ats, Passeig de Llu\'{\i}s Companys, 23, 08010 Barcelona, Spain}\\
$^{25}$ \textit{Institut de F\'{i}sica d’Altes Energies (IFAE), The Barcelona Institute of Science and Technology, Edifici Cn, Campus UAB, 08193, Bellaterra (Barcelona), Spain}\\
$^{26}$ \textit{Department of Physics and Astronomy, University of Waterloo, 200 University Ave W, Waterloo, ON N2L 3G1, Canada}\\
$^{27}$ \textit{Perimeter Institute for Theoretical Physics, 31 Caroline St. North, Waterloo, ON N2L 2Y5, Canada}\\
$^{28}$ \textit{Waterloo Centre for Astrophysics, University of Waterloo, 200 University Ave W, Waterloo, ON N2L 3G1, Canada}\\
$^{29}$ \textit{Instituto de Astrof\'{i}sica de Andaluc\'{i}a (CSIC), Glorieta de la Astronom\'{i}a, s/n, E-18008 Granada, Spain}\\
$^{30}$ \textit{Departament de F\'isica, EEBE, Universitat Polit\`ecnica de Catalunya, c/Eduard Maristany 10, 08930 Barcelona, Spain}\\
$^{31}$ \textit{Department of Physics and Astronomy, Sejong University, 209 Neungdong-ro, Gwangjin-gu, Seoul 05006, Republic of Korea}\\
$^{32}$ \textit{CIEMAT, Avenida Complutense 40, E-28040 Madrid, Spain}\\
$^{33}$ \textit{University of Michigan, 500 S. State Street, Ann Arbor, MI 48109, USA}\\
$^{34}$ \textit{National Astronomical Observatories, Chinese Academy of Sciences, A20 Datun Road, Chaoyang District, Beijing, 100101, P.~R.~China}\\


\bsp	
\label{lastpage}
\end{document}